\documentclass[sigconf,nonacm]{acmart}
\renewcommand\footnotetextcopyrightpermission[1]{}
\usepackage{colortbl}      % \rowcolor / \cellcolor if needed
\usepackage{multirow}
\usepackage{makecell}
\usepackage{tabularx}
\usepackage{array}
\usepackage{enumitem}
\usepackage[most]{tcolorbox}
\usepackage{tikz}
\usetikzlibrary{shapes,arrows,positioning,fit,backgrounds,calc}

\AtBeginDocument{%
  }

\begin{document}

\title{FinSAgent: Corpus-Aligned Multi-Agent RAG Framework for Evidence-Grounded SEC Filing Question Answering}

%% ---------------------------------------------------------------
%% Authors (placeholders; fill in for camera-ready)
%% ---------------------------------------------------------------
% \author{Anonymous Author(s)}
% \affiliation{%
%   \institution{Anonymous Institution}
%   \city{}
%   \country{}
% }
% \email{anon@example.com}

\author{Jijun Chi\textsuperscript{3}, Zhenghan Tai\textsuperscript{1,3}, Hanwei Wu\textsuperscript{1,12}, Tung Sum Thomas Kwok\textsuperscript{1,4}, Hailin He\textsuperscript{1}, Zixing Liao\textsuperscript{1}, Bohuai Xiao\textsuperscript{1}, Chaolong Jiang\textsuperscript{1}, Jianliang Lei\textsuperscript{1}, Jerry Huang\textsuperscript{7,9}, Peng Lu\textsuperscript{7}, Muzhi Li\textsuperscript{5}, Liheng Ma\textsuperscript{1,2,9}, Yihong Wu\textsuperscript{7}, Sicheng Lyu\textsuperscript{1,2,9}, Jingrui Tian\textsuperscript{2}, Yihan Li\textsuperscript{8}, Yanzhang Ma\textsuperscript{1,11}, Sizhe Guan\textsuperscript{1,12}, Dingtao Hu\textsuperscript{2}, Yufei Cui\textsuperscript{2}, \\ Ling Zhou\textsuperscript{10}, Lei Ding\textsuperscript{1,6}, Xinyu Wang\textsuperscript{1,2}}
\affiliation{
  \institution{\textsuperscript{1}SimpleWay.AI\quad \textsuperscript{2}McGill University\quad
\textsuperscript{3}University of Toronto\quad \textsuperscript{4}University of California, Los Angeles\\
\textsuperscript{5}The Chinese University of Hong Kong\quad
\textsuperscript{6}University of Manitoba\quad
\textsuperscript{7}Universit\'e de Montr\'eal\quad \textsuperscript{8}Boston University\\
\textsuperscript{9}Mila - Quebec AI Institute\quad \textsuperscript{10}CG Matrix Technology Limited\quad \textsuperscript{11}Lakehead University\quad \textsuperscript{12}McMaster University }
  \country{}
}
\email{ferris.chi@alumni.utoronto.ca, xinyu.wang5@mail.mcgill.ca}

\renewcommand{\shortauthors}{Chi et al.}

\begin{abstract}
Financial question answering over U.S.\ Securities and Exchange Commission (SEC) filings requires retrieving and synthesizing heterogeneous evidence dispersed across long, standardized, and highly redundant disclosures.
Existing retrieval-augmented and multi-agent systems typically derive retrieval queries directly from the user's question and rank candidates by semantic similarity.
Together, these choices create \emph{prior--corpus misalignment}: a mismatch between model priors and the target filings' structure, terminology, and evidence standards.
As a result, query generation misses corpus-specific evidence, while semantic reranking favors topically similar but evidentially invalid ``false-positive'' chunks.
We propose \textsc{FinSAgent}, an evidence-grounded multi-agent framework that reframes SEC filing QA as corpus-aligned retrieval planning and corrects both ends with a single principle: inject corpus-side conditioning wherever model priors would otherwise dominate.
\textsc{FinSAgent} combines (1) role-specialized agents anchored to the mandated 10-K item structure, (2) database-aware query decomposition that conditions each agent's sub-queries on a lightweight, summary-level view of the local corpus, and (3) multi-path retrieval with a learned feature-gated reranker that separates evidential validity from semantic similarity.
Across five offline financial QA benchmarks, \textsc{FinSAgent} improves retrieval coverage and answer correctness over strong single-agent and multi-agent baselines; in a three-arm randomized online experiment with 1,000 anonymous user ratings, it also receives higher scores than baselines.
\end{abstract}

%% ---------------------------------------------------------------
%% CCS + keywords
%% ---------------------------------------------------------------



\maketitle

\section{Introduction}
\label{sec:intro}

Financial question answering (QA) over U.S.\ Securities and Exchange Commission (SEC) filings asks a system to answer analyst-style questions from annual and quarterly corporate disclosures such as 10-Ks and 10-Qs.
The task matters because filings are among the primary public sources for understanding a firm's operations, risks, strategy, and financial condition.
In practice, users rarely consult filings to recover an isolated fact.
They ask whether a company's expansion appears sustainable, which risks are most material to a strategic shift, or how management's narrative aligns with reported numbers.
Answering such questions requires more than fluent generation: it requires locating the right evidence in long, structured documents and synthesizing responses that remain grounded in disclosed sources~\citep{finqa2021,convfinqa2022,financebench2023,docfinqa2024,secque2025,jiang2026fin,zhu2024tat,gupta2024systematic,yepes2024financial,r2r}.

We focus on \emph{evidence-grounded} financial QA over a filing database, where answering a question often requires combining numerical and narrative evidence scattered across business descriptions, management discussion, risk factors, legal proceedings, footnotes, and financial tables, while near-duplicate boilerplate recurs across issuers and years.
The task is therefore not merely long-context prompting, but requires both corpus-aligned retrieval planning and validity-aware evidence selection over a large, structured, and highly redundant corpus.
Recent work has strengthened individual components through domain-adapted language models \citep{BloombergGPT,FinGPT,XuanYuan}, retrieval-centric preprocessing, indexing, and reranking~\citep{FinSage,veritasfi2025}, and role-specialized multi-agent reasoning~\citep{findebate,finteam2025,nguyen2025ma,zhu2025role}.
Yet it leaves a central question unresolved: \emph{how should a system discover and select valid evidence for complex filing questions?}
% Existing systems typically follow a single query trajectory or decompose the user question using model priors alone, then rank candidates primarily by semantic similarity.

% We focus on \emph{evidence-grounded} financial QA over a filing database. This is
% hard because the supporting evidence is typically long-range, heterogeneous, and
% distributed across sections: relevant support may lie in business descriptions,
% management discussion, risk factors, legal proceedings, footnotes, and financial
% tables, and a single question may require fusing numerical and narrative evidence
% from several of these parts. Many filing questions are also inherently
% multi-faceted, involving company fundamentals, market interpretation, and legal
% or regulatory disclosure at once. SEC filing QA is therefore not a long-context
% prompting problem but a problem of \emph{planning and validating evidence search}
% over a large, structured, and highly redundant corpus.

\begin{figure}[t]
    \centering
    \includegraphics[width=0.95\linewidth]{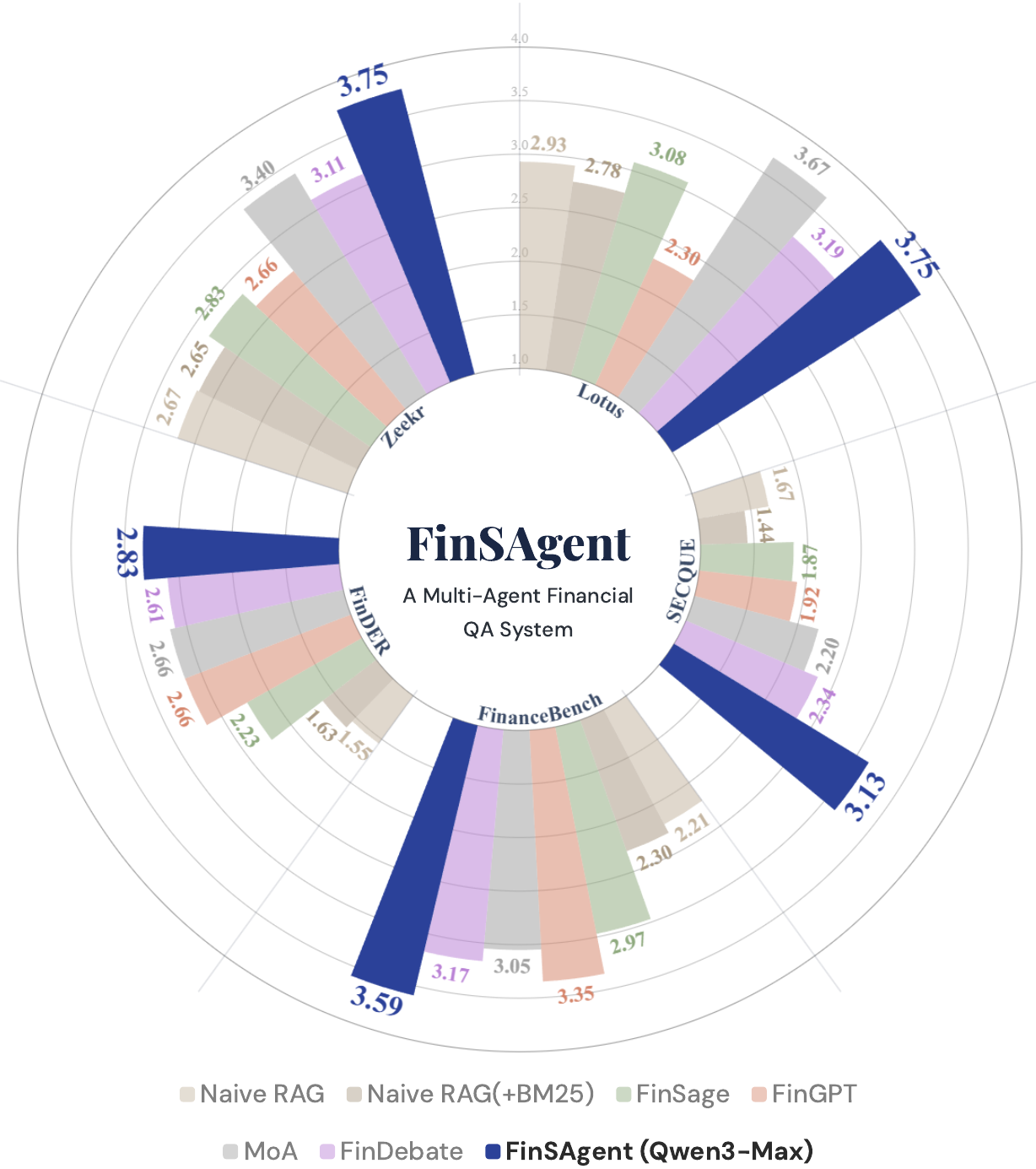}
    \caption{Answer correctness across five financial QA benchmarks.
    \textsc{FinSAgent} attains the highest correctness on every benchmark,
    outperforming single-agent RAG pipelines and multi-agent baselines under a
    matched evidence budget.}
    \label{fig:RadialChart}
    \vspace{-0.88cm}
\end{figure}

% Recent work has improved many ingredients for this task. Domain-adapted language
% models strengthen general financial language understanding
% \citep{BloombergGPT,FinGPT,XuanYuan}; retrieval-centric systems improve grounding
% through document preprocessing, metadata-aware indexing, multi-path retrieval,
% and domain-specialized reranking~\citep{FinSage,veritasfi2025}; and multi-agent
% methods show that role specialization helps organize complex reasoning in
% financial and document-centric settings~\citep{findebate,finteam2025,nguyen2025ma,zhu2025role}.
% On their own, however, these advances leave a central question unresolved:
% \emph{how should a system plan evidence retrieval for complex filing questions?}
% Existing systems overwhelmingly follow a single query trajectory, or decompose
% queries directly from the user question, and then rank candidates by semantic
% similarity.

\paragraph{The core problem.}
We use \emph{prior--corpus misalignment} to describe the mismatch between model priors and the target filing corpus's structure, terminology, and criteria for valid evidence. This mismatch affects both query formation and evidence selection (Figure~\ref{fig:framework}).
% For example, for the question ``What are Zeekr's main risks?'', the FinSage~\citep{FinSage} baseline fills its top-8 evidence budget with recurring PRC listing boilerplate while omitting company-specific evidence on sales concentration and European expansion. We analyze this failure quantitatively in \S\ref{sec:case-study}.
At the \textbf{front end}, prior-driven query decomposition produces sub-queries that are semantically plausible yet misaligned with how evidence is actually written and organized in the filings, so genuinely disclosed material is never retrieved.
At the \textbf{back end}, a semantic reranker rewards chunks that match the same priors, so semantic relevance separates from evidential validity: near-verbatim boilerplate (e.g., generic risk-factor or compliance language that recurs across issuers) scores highly yet is invalid for the specific company, crowding out the true disclosure.
Effective retrieval therefore requires corpus alignment at both stages: retrieval planning determines whether relevant evidence enters the candidate set, while evidence selection determines whether it survives the final evidence budget.

\paragraph{FinSAgent.}
To address both stages, we propose \textsc{FinSAgent} (Figure~\ref{fig:framework}), an evidence-grounded multi-agent framework built around this problem.
The central design principle is to \emph{inject corpus-side conditioning wherever model priors would otherwise dominate}, so that evidence exploration is broadened while staying aligned with the local corpus and conservative about evidence quality.
\textsc{FinSAgent} instantiates three coupled mechanisms:
(1)~\textbf{role-specialized parallel agents}, anchored to the mandated 10-K item structure, that separate heterogeneous evidence needs (general, quantitative, market, legal, and company) before retrieval to broaden coverage;
(2)~\textbf{database-aware query decomposition}, which conditions each agent's sub-query generation on a lightweight, summary-level view of the local filing database rather than the user question alone; 
and (3)~\textbf{multi-path retrieval with feature-gated reranking}, in which a learned gate over non-semantic features down-weights confident false positives that a semantic reranker cannot separate.
The first two components support corpus-aligned retrieval planning, while the third performs validity-aware evidence selection.

% ! This paragraph argues db-aware decomp and feature-gate rerank effect in different stages, and feature gate is non-trivial. NOt needed in introduction.

% The two corrective mechanisms are not two verifiers, nor a re-combination of
% off-the-shelf parts: they are two responses to a single cause. Database-aware
% decomposition acts \emph{before} retrieval as information supplementation---it
% can recover evidence that would otherwise never surface, a gain no post-hoc
% verifier can produce---while feature-gated reranking acts \emph{after} retrieval
% and models validity as separable from similarity. We support this claim
% empirically: a mechanistic SHAP analysis (\S\ref{sec:gating-ablation}) shows that
% a small set of non-semantic features carries the validity signal that the
% reranker systematically misses, turning a design choice into an interpretable
% finding that should transfer to any corpus with prior--corpus divergence and
% redundancy-induced false positives.

Our contributions are:
\begin{itemize}
    % \item We \textbf{reframe} SEC filing QA as corpus-aligned retrieval planning and identify \emph{prior--corpus misalignment} as a single cause that surfaces symmetrically at query formation and evidence selection (\S\ref{sec:motivation}), characterizing why filings are structurally distinct from generic long-document RAG.
    \item We frame SEC filing QA as requiring alignment at both retrieval planning stage and evidence selection stage, and introduce \emph{prior-corpus misalignment} as a unifying lens for failures at both stages.
    % \item We propose \textbf{database-aware query decomposition}, a testable corpus-conditioned decomposition mechanism that generates agent-specific sub-queries from a lightweight local view of the filing database (\S\ref{sec:dad}).
    % \item We identify and learn to correct the \textbf{semantic--validity decoupling} in filing retrieval via a feature-gated reranker, and show with SHAP that non-semantic features carry the missing validity signal (\S\ref{sec:retrieval}, \S\ref{sec:gating-ablation}).
    \item We introduce \textsc{FinSAgent}, which combines role-specialized, database-aware retrieval planning with multi-path retrieval and feature-gated evidence selection for candidate selection (\S\ref{sec:dad}, \S\ref{sec:retrieval}).
    \item Across \textbf{five benchmarks}, \textsc{FinSAgent} improves retrieval coverage and answer correctness. Tests under production scenario further supports these results: across $\sim$1,400 anonymous user ratings, \textsc{FinSAgent} receives higher scores than baselines (\S\ref{sec:experiments}).
\end{itemize}

\section{Related Work}
\label{sec:related}

\paragraph{Financial foundation models.}
The linguistic and numerical complexity of financial corpora has driven a line of
domain-specialized language models~\citep{li2025time}. Early efforts such as
BloombergGPT~\citep{BloombergGPT} and open-source frameworks like
FinGPT~\citep{FinGPT} and XuanYuan~\citep{XuanYuan} demonstrate strong performance
on domain tasks such as sentiment analysis and entity extraction. Recent work
explores multimodality by post-training on textual and visual financial
data~\citep{wang2023finvisgptmultimodallargelanguage,bhatia-etal-2024-fintral}:
TAT-LLM~\citep{TAT-LLM} targets discrete reasoning over hybrid tabular--textual
data, and Open-FinLLMs~\citep{huang2025openfinllmsopenmultimodallarge} extends
coverage across text, tables~\citep{xing2026tabledart,kwok2026enhancingtableqaverifiablereasoning},
time series~\citep{liu-etal-2025-picture}, and charts~\citep{10.1145/3677052.3698696}.
These models encode deep domain knowledge but operate as standalone
generators without cross-sourced reasoning~\citep{ctx64989054940006531} across
heterogeneous evidence, and thus struggle with the multi-step, longitudinal
reasoning that long-form SEC filings demand.

\paragraph{Agentic and retrieval-augmented systems.}
Retrieval-augmented generation (RAG) grounds model outputs in verifiable
evidence and has become the prevalent approach to financial
QA~\citep{zhao-etal-2025-finragbench,10841711}. FinSage~\citep{FinSage} adopts a
deterministic ``extract-and-summarize'' workflow built on multi-path retrieval
and a domain-specialized reranker, but its static pipeline cannot dynamically
align queries with heterogeneous filing sections~\citep{10.1145/3746252.3761643}.
A parallel line pivots to multi-agent frameworks: FinDebate~\citep{findebate}
combines role-specialized agents with a safety-constrained debate protocol to
curb model overconfidence~\citep{yan2026learn}, FinSight~\citep{FinSight}
coordinates specialists for data collection, context analysis, and report
generation through a shared programmable variable space, and mixture-of-agents
designs aggregate independent specialist drafts~\citep{wang2024mixture,wu2024autogen,marag2025,mdocagent2025}.
On the retrieval-planning side, query decomposition
methods~\citep{qdrag2026,petcu2025query,wu2025advancing,zhang2024exploring}
improve coverage by splitting complex questions into sub-queries, but remain
primarily \emph{question-driven}: they refine or expand the original query without
conditioning on the local structure and terminology of the target corpus, a
limitation shared with classical pseudo-relevance
feedback~\citep{pseudo,chen-etal-2024-analyze}. Uncertainty-aware
retrieval~\citep{bayesianrag,bayesianrag2026} highlights the importance of
evidence reliability but typically relies on expensive stochastic inference and
is under-explored for structured filing QA.

\textsc{FinSAgent} differs from these systems along the axis that matters for filing QA: it treats retrieval as a corpus-aligned planning problem rather than a question-conditioned function.
Concretely, it introduces database-aware decomposition to condition sub-queries on a lightweight corpus view, and augments multi-path retrieval with a learned feature gate that models evidential validity as separable from semantic similarity.

\section{The FinSAgent Framework}
\label{sec:method}

\subsection{Problem Formulation and Overview}
\label{sec:overview}

We formulate SEC filing QA as evidence-grounded generation over a filing database $\mathcal{D}$.
Given a user question $q$, the goal is to retrieve supporting evidence $E \subset \mathcal{D}$ and generate an answer $y$ grounded in $E$.
Unlike general multi-hop QA, which typically requires a sequential reasoning chain across entities, SEC filing questions demand \emph{evidence aggregation across heterogeneous perspectives} that are scattered over fixed sections and mixed table--prose formats.

\begin{figure*}[t]
    \centering
    \makebox[\textwidth][c]{\includegraphics[width=0.8\textwidth]{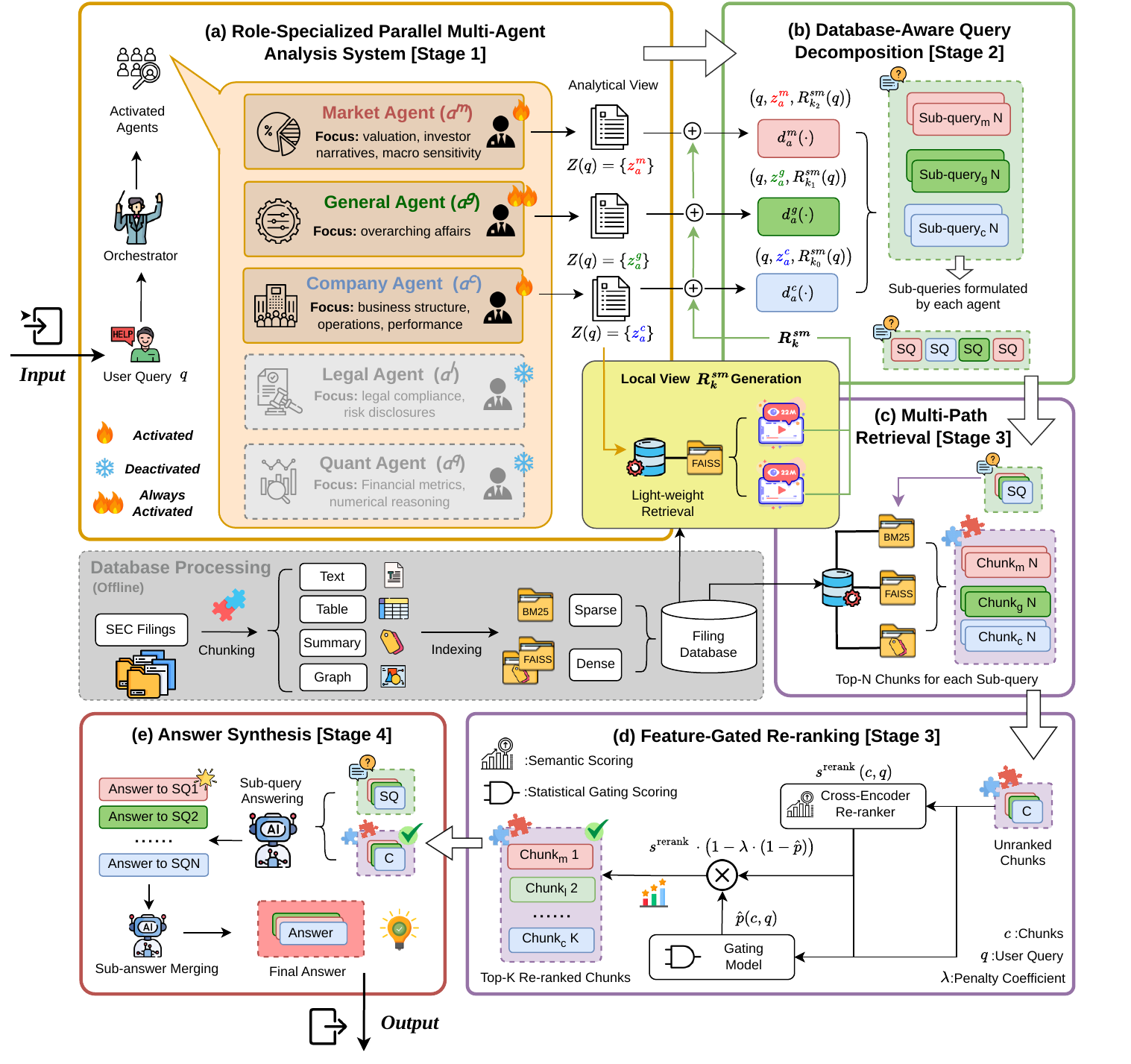}}
    \caption{Overall framework of \textsc{FinSAgent}. Role-specialized agents
    first construct complementary analytical views. Database-aware decomposition
    then conditions each agent's sub-queries on a lightweight summary-level view
    of the local filing database (correcting the front-end query gap), and
    multi-path retrieval with a feature-gated reranker selects reliable evidence
    (correcting the back-end validity gap), before evidence aggregation and final
    answer synthesis.}
    \label{fig:framework}
\end{figure*}

Following the two-ended view of prior--corpus misalignment introduced in \S\ref{sec:intro}, \textsc{FinSAgent} is a three-stage pipeline (Figure~\ref{fig:framework}):
(1)~a \textbf{role-specialized multi-agent module} constructs complementary general, quantitative, market, legal, and company views of the question and broadens coverage (\S\ref{sec:multiagent});
(2)~a \textbf{database-aware decomposition module} generates agent-specific sub-queries conditioned on a lightweight, summary-level view of the local corpus, correcting the front-end gap (\S\ref{sec:dad}); and
(3)~a \textbf{multi-path retrieval module with feature-gated reranking} selects reliable evidence, correcting the back-end gap (\S\ref{sec:retrieval}).
An orchestrator routes $q$ to the relevant agents and aggregates their outputs
into the final grounded answer. An optional conversational-memory extension for
multi-turn analysis is described in Appendix~\ref{app:memory}.

\subsection{Role-Specialized Parallel Multi-Agent System}
\label{sec:multiagent}

SEC filing QA requires evidence distributed across multiple parts of a filing,
which a single search trajectory captures poorly. To improve coverage,
\textsc{FinSAgent} first decomposes the question into complementary
evidence-seeking views using a role-specialized parallel multi-agent module.

Let $\mathcal{A}=\{a^{g}, a^{q}, a^{m}, a^{l}, a^{c}\}$ denote agents for
general, quantitative, market, legal, and company evidence. Each agent
$a\in\mathcal{A}$ carries a role policy $\pi_a=(\rho_a,\Omega_a,\phi_a)$, where
$\rho_a$ specifies its responsibility, $\Omega_a$ its retrieval scope, and
$\phi_a$ its analytical focus. Given $q$, each agent produces a role-conditioned
analytical view $z_a = g_a(q;\pi_a)$, where $g_a(\cdot)$ is the prompting function
induced by $\pi_a$ and $z_a$ is a short natural-language interpretation of $q$
that guides downstream decomposition and retrieval. All agents operate in
\emph{parallel} on the same question, disentangling heterogeneous evidence needs
before retrieval and reducing the risk that one search path overlooks relevant
support. Role prompts are given in Appendix~\ref{app:multi-agent-prompt}.

\paragraph{Why these roles.}
The agent roles are derived from recurring evidence categories reflected in the standardized 10-K item structure\footnote{\url{https://www.sec.gov/files/form10-k.pdf}} and expert-validated (Table~\ref{tab:roles}).
% Roles that map to items keep evidence needs grounded in the corpus; candidate roles that are \emph{not} item-anchored (e.g., trend, sentiment, or ESG analysts) tend to reduce to question-side decompositions---exactly the prior-driven front-end failure mode we target in \S\ref{sec:misalignment}---or overlap existing roles at added token cost.
% The item structure is thus a natural saturation point: fewer roles lose exclusive evidence (\S\ref{sec:retrieval-results} shows even the lowest-activation Legal role retrieves ground-truth chunks no other role reaches), while more roles add cost without new item coverage.
As shown in Table~\ref{tab:roles}, the general, quantitative, market, legal, and company agents cover complementary retrieval scopes across business disclosures, management discussion, risk and legal sections, and financial statements.
This mapping is deliberately created to tie agents' queries to specific filing sections and distinct retrieval scopes.
The general agent remains active for cross-item and definitional questions, while the specialized agents target their corresponding evidence categories.
Our retrieval analysis indicates that role specialization provides complementary coverage(\S\ref{sec:retrieval-results}). We therefore adopt these five roles as a practical trade-off between evidence coverage and inference cost.

\begin{table}[t]
\centering
\small
\caption{Agent roles anchored to the mandated 10-K item structure.}
\label{tab:roles}
\begin{tabular}{@{}lll@{}}
\toprule
\textbf{Agent} & \textbf{Focus} & \textbf{10-K item(s)} \\
\midrule
General      & Cross-item / definitional        & (always on) \\
Quantitative & Financial metrics, numerics      & Item 8, 7A \\
Market       & Industry context, positioning    & Item 7 (MD\&A) \\
Legal        & Legal proceedings, compliance    & Item 1A, 3 \\
Company      & Business, operations, segments   & Item 1, 1B--2 \\
\bottomrule
\end{tabular}
\end{table}

\subsection{Database-Aware Query Decomposition}
\label{sec:dad}

After obtaining the role-conditioned views $\{z_a\}_{a\in\mathcal{A}}$, each agent
decomposes the question into retrieval-ready sub-queries. Role specialization
alone does not make this reliable: an LLM tends to generate sub-queries from its
internalized priors, which differ substantially from the target corpus and lead
to weak or missed retrievals.

We therefore ground decomposition by exposing each agent to a coarse-grained
local view of the filing database. During offline preprocessing we generate
compact summaries for contiguous spans of filing chunks and index them. At query
time, a lightweight FAISS-based retriever~\citep{douze2025faiss} maps $q$ to the
top-$k_0$ most relevant \emph{section-level} summaries, forming the local
database view $R^{\mathrm{sv}}_{k_0}(q)$. This step is intentionally lightweight:
it does not answer the question, but exposes which filing regions are likely
task-relevant and surfaces corpus-specific terminology. Conditioned on this view,
each agent performs role-specific decomposition:
\[
\mathcal{Q}_a = d_a\!\left(q,\; z_a,\; R^{\mathrm{sv}}_{k_0}(q)\right)
= \{q_{a,1}, \dots, q_{a,n_a}\},
\]
where $d_a$ is a prompted LLM call that receives $q$, the analytical view $z_a$, and the retrieved summaries as structured input and emits a list of role-specific sub-queries.

\paragraph{Dynamic corpus conditioning.}
The agent prompt templates are fixed, but the corpus view $R^{\mathrm{sv}}_{k_0}(q)$ is retrieved anew for each question.
Unlike a static terminology prompt or a hand-written overview, this query-dependent view exposes the filing regions and corpus-specific terms relevant to the current question.
The resulting sub-queries can therefore reflect the filing's section headings, segment names, and disclosure terms through a lightweight summary lookup, without encoding the entire database in the prompt.
% This matters because filings contain issuer-specific section names, segment labels, and disclosure terms that a generic terminology prompt cannot anticipate.
% A static issuer specification alternative fails in two ways: enumerating the whole database becomes infeasible and noisy as the corpus grows, whereas a hand-written overview cannot identify \emph{which} local section, segment, or term is relevant to the current question.
% This lightweight summary lookup is low-cost yet dynamic, which is why we view database-aware decomposition as \emph{corpus-grounded} decomposition rather than mere terminology adaptation.

\subsection{Multi-Path Retrieval and Feature-Gated Reranking}
\label{sec:retrieval}

Given the sub-query sets $\{\mathcal{Q}_a\}_{a\in\mathcal{A}}$, \textsc{FinSAgent}
retrieves candidates within each agent's scope and filters them through a
feature-gated reranker: multi-path retrieval for broad coverage, then reranking
for evidence reliability.

\subsubsection{Multi-Path Evidence Retrieval}
A single retrieval signal is often insufficient, since relevant evidence may
appear under exact lexical matches, semantic paraphrases, or coarse section-level
summaries. For each agent $a$ and sub-query $q_{a,j}\in\mathcal{Q}_a$, we
construct a candidate set $C_{a,j}$ as the union of the top-$k$ results from three
complementary paths restricted to the agent's scope $\Omega_a$:
\[
C_{a,j}
=
R^{\mathrm{sp}}_k(q_{a,j};\Omega_a)
\;\cup\;
R^{\mathrm{de}}_k(q_{a,j};\Omega_a)
\;\cup\;
R^{\mathrm{sm}}_k(q_{a,j};\Omega_a),
\]
where $R^{\mathrm{sp}}$ is sparse retrieval via BM25, $R^{\mathrm{de}}$ is dense
retrieval via cosine similarity over chunk embeddings, and $R^{\mathrm{sm}}$ is
summary-path retrieval over precomputed section-level summary embeddings. A
dedicated table branch (\S\ref{app:preprocess}) supplies structured tabular
evidence. Formal definitions are given in Appendix~\ref{app:retrieval}.

\subsubsection{Feature-Gated Evidence Reranking}
Because the candidate set $C_{a,j}$ is constructed for high recall, it can include topically relevant but evidentially invalid chunks that a cross-encoder reranker cannot reliably distinguish.
\textsc{FinSAgent} therefore augments its semantic score with a learned feature gate.
% The candidate set $C_{a,j}$ is deliberately over-inclusive for high recall. In SEC filings, many retrieved chunks are topically related but not supportive: adjacent sections reuse language, boilerplate recurs across filings, and generic statements resemble true evidence.
% A cross-encoder reranker improves semantic ordering, but similarity alone cannot separate valid evidence from confident false positives (\S\ref{sec:misalignment}).
% \textsc{FinSAgent} therefore augments the reranker with a learned feature gate.

For each candidate $c \in C_{a,j}$ we compute a base reranker score
$s^{\mathrm{rr}}(c, q_{a,j})$ and a lightweight feature vector
\vspace{-0.1cm}
\[
\mathbf{x}(c, q_{a,j}) =
\big[
\mathbf{s},\;
\mathbf{r},\;
\mathbf{f}_{\mathrm{query}},\;
\mathbf{f}_{\mathrm{chunk}},\;
\mathbf{f}_{\mathrm{overlap}},\;
s^{\mathrm{rr}}
\big],
\]
\vspace{-0.1cm}
where $\mathbf{s}$ collects the three path scores, $\mathbf{r}$ records
path-specific rank and provenance (e.g., how many paths surfaced the chunk),
$\mathbf{f}_{\mathrm{query}}$ captures query-level statistics,
$\mathbf{f}_{\mathrm{chunk}}$ chunk-level metadata, and
$\mathbf{f}_{\mathrm{overlap}}$ lexical-overlap features. Critically, these are
\emph{non-semantic} signals the cross-encoder does not model. An offline-trained
LightGBM model~\citep{ke2017lightgbm} $\mathcal{M}$ maps them to a validity
estimate $\hat{p}(c, q_{a,j}) = \mathcal{M}(\mathbf{x}) \in [0,1]$; we choose a
gradient-boosted tree for its low inference overhead and interpretability. The
estimate combines with the reranker score through a multiplicative gate:
\[
\tilde{s}(c, q_{a,j})
=
s^{\mathrm{rr}}(c, q_{a,j})
\left(1 - \lambda_a \big(1-\hat{p}(c, q_{a,j})\big)\right),
\]
where $\lambda_a \in [0,1]$ is the per-agent gating strength tuned on a
validation split; performance is stable across a broad range
(\S\ref{sec:gating-ablation}). The multiplicative form scales the penalty with
the reranker's own confidence, so a high-scoring chunk judged risky is demoted
more than a low-scoring one---confident false positives are the more damaging
case. The final evidence set is
$E_{a,j} = \operatorname{TopK}^{k_e}_{c \in C_{a,j}} \tilde{s}(c, q_{a,j})$.
Each agent then drafts a grounded response over $E_{a,j}$, and the orchestrator
concatenates the retrieved evidence and drafts to synthesize the final answer to
$q$.

\paragraph{Gating model training.}
We construct the training set exclusively from an in-distribution training split to prevent leakage.
Using targeted hard-negative mining, an LLM generates questions from random chunks; these are run through the full retrieval and reranking pipeline; within the top-$k$ results, chunks matching the chosen ground truth are labeled positive and the rest serve as competitive hard negatives, downsampled to an approximate $20\%$ positive rate.
% This teaches the gate to distinguish true evidence from competitive false positives within the target distribution.
Full details, the 31-feature specification, and a SHAP analysis are in Appendix~\ref{app:gating}.

\section{Experiments and Results}
\label{sec:experiments}

We evaluate whether \textsc{FinSAgent} improves evidence-grounded filing QA and,
crucially, whether its gains come from \emph{corpus alignment} rather than from
added retrieval budget or extra deliberation. We report end-to-end quality
(\S\ref{sec:e2e}), retrieval coverage and component ablations
(\S\ref{sec:retrieval-results}), a comparison
to full-document long-context prompting (\S\ref{sec:longcontext}), evaluation-validity
evidence including a blind human study and RAGAS noise sensitivity
(\S\ref{sec:validity}), system overhead and per-phase latency
(\S\ref{sec:overhead}), a case study (\S\ref{sec:case-study}), and an error
analysis (\S\ref{sec:error-analysis}). Appendices~\ref{app:fairness} and~\ref{app:gating} report a matched-budget fairness study and a gating-strength ablation with SHAP attribution.

\subsection{Experimental Setup}
\label{sec:setup}

\paragraph{Datasets.}
We evaluate on five financial QA datasets spanning public benchmarks and
proprietary in-house collections. The public benchmarks are the open-source
version of \texttt{FinanceBench}~\citep{financebench2023}, the
\texttt{FinDER} dataset~\citep{choi2025finderfinancialdatasetquestion}, and a
100-question stratified subset of \texttt{SECQUE}~\citep{secque2025} that
preserves SECQUE's native question-type distribution (Appendix~\ref{app:preprocess}). The two proprietary datasets, \texttt{Lotus} and
\texttt{Zeekr}, are drawn from real-world business data for two automotive brands
and contain multi-entity QA pairs with annotated relevance labels. Corpus
construction (MinerU parsing, modality-tagged chunking, table/figure branches,
near-duplicate removal, coreference resolution, section-level metadata, and dense
BGE-M3 plus sparse BM25 indices) is detailed in Appendix~\ref{app:preprocess}.

\paragraph{Baselines and backbones.}
We compare against traditional and agentic systems: Naive RAG (dense-only and
+BM25 variants), FinSage~\citep{FinSage} (single agent + multi-path retrieval),
FinGPT~\citep{FinGPT} (single-agent tool-augmented reasoner), and the multi-agent
systems MoA~\citep{wang2024mixture} and FinDebate~\citep{findebate}. To avoid
confounding gains with backbone choice, all baselines use the same generator
backbone as \textsc{FinSAgent} within each comparison (DeepSeek-v3.1-Terminus),
and we additionally report \textsc{FinSAgent} under Kimi-K2.5 and Qwen3-Max to
show that the gains are robust to backbone choice. The base reranker is bge-reranker-v2-gemma, and the learned gate is
a LightGBM model over the 31 features in Appendix~\ref{app:gating}. Baseline
implementation details are in Appendix~\ref{app:baselines}.

\paragraph{Metrics and judging protocol.}
We report answer \textbf{Correctness} together with five Likert-scale
(1--5)~\citep{likert1932technique} dimensions adopted from
Fin-RATE~\citep{jiang2026fin}: \textbf{Information Coverage},
\textbf{Reasoning Chain}, \textbf{Factual Consistency}, \textbf{Clarity of
Expression}, and \textbf{Analytical Depth}. For retrieval we report Macro-Recall
against annotated relevance labels. This suite subsumes the metrics commonly used
in this space: our Correctness is a finer-grained (1--5) version of SECQUE-Judge's
0/1/2 correctness and of RAGAS answer accuracy; Factual Consistency corresponds to
RAGAS faithfulness; and Macro-Recall corresponds to RAGAS context recall. We add
RAGAS noise sensitivity in \S\ref{sec:validity}.

Judging is \textbf{blind and order-randomized}: each candidate answer is stripped
of any system identifier and presented in randomized order, and the judge receives
only the question, the reference answer, and the candidate answer. Because all
systems in a comparison share the same backbone that also serves as the judge, any
self-preference of the judge applies equally to every system and cannot favor
\textsc{FinSAgent}; \S\ref{sec:validity} provides an independent human check.

\begin{table*}[tp]
\centering
\footnotesize
\caption{End-to-end financial QA performance. \textbf{Corr.}=Correctness,
\textbf{Info.}=Information Coverage, \textbf{Reas.}=Reasoning Chain,
\textbf{Fact.}=Factual Consistency, \textbf{Clar.}=Clarity, \textbf{Depth}=Analytical
Depth (all higher is better). \textbf{Best} in bold, \underline{second-best}
underlined, per dataset. *On FinanceBench we additionally include a full-document
long-context baseline (LC) with frontier model, which reads the entire 10-K with no retrieval
(\S\ref{sec:longcontext}).}
\label{tab:end2end}
\setlength{\tabcolsep}{4.5pt}
\renewcommand{\arraystretch}{0.95}
\begin{tabular}{@{}lll c ccccc@{}}
\toprule
\textbf{Dataset} & \textbf{Type} & \textbf{Method} & \textbf{Corr.} & \textbf{Info.} & \textbf{Reas.} & \textbf{Fact.} & \textbf{Clar.} & \textbf{Depth} \\
\midrule
\multirow{9}{*}{\textbf{Lotus}}
& Single & Naive RAG            & 2.93 & 2.98 & 3.45 & 3.68 & 4.50 & 3.05 \\
& Single & Naive RAG (+BM25)    & 2.78 & 2.83 & 3.37 & 3.50 & 4.39 & 2.94 \\
& Single & FinSage              & 3.08 & 3.26 & 3.75 & 3.99 & 4.53 & 3.48 \\
& LC* & Long-context (full 10-K) & 0 & 0 & 0 & 0 & 0 & 0 \\
\cmidrule(lr){2-9}
& Web    & FinGPT               & 2.30 & 2.82 & 3.13 & 3.49 & 4.43 & 3.07 \\
\cmidrule(lr){2-9}
& Multi  & MoA                  & \underline{3.67} & \underline{4.03} & \underline{4.14} & 4.11 & 4.62 & \underline{4.21} \\
& Multi  & FinDebate            & 3.19 & 3.58 & 3.90 & 3.89 & \textbf{4.69} & 3.96 \\
\cmidrule(lr){2-9}
& Multi  & \textsc{FinSAgent} (DS-v3.1)  & 3.58 & 3.79 & 3.91 & \textbf{4.16} & 4.58 & 3.83 \\
& Multi  & \textsc{FinSAgent} (Kimi-K2.5) & 3.57 & 3.98 & 3.94 & 3.87 & 4.42 & 4.04 \\
& Multi  & \textsc{FinSAgent} (Qwen3-Max) & \textbf{3.75} & \textbf{4.20} & \textbf{4.28} & \underline{4.12} & \underline{4.63} & \textbf{4.32} \\
\midrule
\multirow{9}{*}{\textbf{SECQUE}}
& Single & Naive RAG            & 1.67 & 1.68 & 1.90 & 2.20 & 4.24 & 1.70 \\
& Single & Naive RAG (+BM25)    & 1.44 & 1.63 & 1.81 & 1.91 & 4.28 & 1.69 \\
& Single & FinSage              & 1.87 & 2.25 & 2.27 & 2.51 & 4.08 & 2.13 \\
& LC* & Long-context (full 10-K) & 0 & 0 & 0 & 0 & 0 & 0 \\
\cmidrule(lr){2-9}
& Web    & FinGPT               & 1.92 & 2.49 & 2.81 & 3.02 & 4.06 & 2.70 \\
\cmidrule(lr){2-9}
& Multi  & MoA                  & 2.20 & 2.81 & 3.07 & 2.66 & 4.28 & 3.05 \\
& Multi  & FinDebate            & 2.34 & 2.88 & 3.26 & 2.73 & 4.36 & 3.17 \\
\cmidrule(lr){2-9}
& Multi  & \textsc{FinSAgent} (DS-v3.1)  & \underline{2.96} & \underline{3.38} & \underline{3.39} & \underline{3.35} & \underline{4.44} & \underline{3.23} \\
& Multi  & \textsc{FinSAgent} (Kimi-K2.5) & 2.64 & 3.17 & 3.31 & 2.90 & 4.19 & 3.19 \\
& Multi  & \textsc{FinSAgent} (Qwen3-Max) & \textbf{3.13} & \textbf{3.81} & \textbf{3.87} & \textbf{3.56} & \textbf{4.60} & \textbf{3.82} \\
\midrule
\multirow{10}{*}{\textbf{FinanceBench}}
& Single & Naive RAG            & 2.21 & 2.36 & 2.52 & 2.76 & 4.54 & 2.40 \\
& Single & Naive RAG (+BM25)    & 2.30 & 2.28 & 2.47 & 2.96 & 4.54 & 2.34 \\
& Single & FinSage              & 2.97 & 3.27 & 3.34 & 3.26 & 4.60 & 3.23 \\
& LC* & Long-context (full 10-K) & 2.98 & 3.29 & 3.52 & 3.37 & 4.60 & 3.42 \\
\cmidrule(lr){2-9}
& Web    & FinGPT               & \underline{3.35} & \textbf{4.10} & \textbf{4.35} & \underline{3.59} & \textbf{4.90} & \textbf{4.21} \\
\cmidrule(lr){2-9}
& Multi  & MoA                  & 3.05 & 3.43 & 3.79 & 3.36 & 4.55 & 3.64 \\
& Multi  & FinDebate            & 3.17 & 3.57 & 4.01 & 3.56 & 4.68 & 3.92 \\
\cmidrule(lr){2-9}
& Multi  & \textsc{FinSAgent} (DS-v3.1)  & 3.32 & 3.70 & 3.77 & 3.52 & 4.61 & 3.71 \\
& Multi  & \textsc{FinSAgent} (Kimi-K2.5) & 3.29 & 3.59 & 3.83 & 3.50 & 4.56 & 3.74 \\
& Multi  & \textsc{FinSAgent} (Qwen3-Max) & \textbf{3.59} & \underline{3.93} & \underline{4.12} & \textbf{3.68} & \underline{4.74} & \underline{4.07} \\
\midrule
\multirow{9}{*}{\textbf{FinDER}}
& Single & Naive RAG            & 1.55 & 1.52 & 1.65 & 2.35 & 4.14 & 1.48 \\
& Single & Naive RAG (+BM25)    & 1.63 & 1.56 & 1.69 & 2.46 & 3.97 & 1.51 \\
& Single & FinSage              & 2.23 & 1.93 & 2.13 & 3.08 & 3.99 & 1.89 \\
& LC* & Long-context (full 10-K) & 0 & 0 & 0 & 0 & 0 & 0 \\
\cmidrule(lr){2-9}
& Web    & FinGPT               & 2.66 & 2.56 & 3.04 & 3.32 & \underline{4.35} & 2.96 \\
\cmidrule(lr){2-9}
& Multi  & MoA                  & 2.66 & 2.70 & 3.24 & \textbf{3.58} & 4.25 & 3.10 \\
& Multi  & FinDebate            & 2.61 & 2.68 & \underline{3.32} & 3.25 & \textbf{4.44} & \underline{3.24} \\
\cmidrule(lr){2-9}
& Multi  & \textsc{FinSAgent} (DS-v3.1)  & \underline{2.68} & \underline{2.83} & 3.15 & \underline{3.56} & 4.25 & 3.01 \\
& Multi  & \textsc{FinSAgent} (Kimi-K2.5) & 2.38 & 2.82 & 3.04 & 2.93 & 3.90 & 3.00 \\
& Multi  & \textsc{FinSAgent} (Qwen3-Max) & \textbf{2.83} & \textbf{3.30} & \textbf{3.48} & 3.41 & 4.32 & \textbf{3.52} \\
\midrule
\multirow{9}{*}{\textbf{Zeekr}}
& Single & Naive RAG            & 2.67 & 2.47 & 2.96 & 3.56 & 4.37 & 2.50 \\
& Single & Naive RAG (+BM25)    & 2.65 & 2.45 & 3.02 & 3.58 & 4.41 & 2.56 \\
& Single & FinSage              & 2.83 & 2.76 & 3.32 & 3.96 & 4.47 & 2.84 \\
& LC* & Long-context (full 10-K) & 0 & 0 & 0 & 0 & 0 & 0 \\
\cmidrule(lr){2-9}
& Web    & FinGPT               & 2.66 & 3.27 & 3.69 & 3.26 & \underline{4.61} & 3.60 \\
\cmidrule(lr){2-9}
& Multi  & MoA                  & 3.40 & 3.57 & 4.02 & 3.95 & 4.56 & 3.92 \\
& Multi  & FinDebate            & 3.11 & 3.40 & 3.95 & 3.83 & \textbf{4.65} & 3.92 \\
\cmidrule(lr){2-9}
& Multi  & \textsc{FinSAgent} (DS-v3.1)  & \underline{3.46} & 3.72 & 3.97 & \underline{4.19} & 4.59 & 3.84 \\
& Multi  & \textsc{FinSAgent} (Kimi-K2.5) & 3.43 & \underline{3.76} & \underline{4.08} & 3.96 & 4.42 & \underline{4.04} \\
& Multi  & \textsc{FinSAgent} (Qwen3-Max) & \textbf{3.75} & \textbf{4.16} & \textbf{4.20} & \textbf{4.20} & \textbf{4.65} & \textbf{4.18} \\
\bottomrule
\end{tabular}
\end{table*}

\subsection{End-to-End QA Performance}
\label{sec:e2e}

Table~\ref{tab:end2end} reports end-to-end quality. Two patterns stand out.
First, the multi-agent paradigm broadly outperforms single-agent baselines,
confirming that collaborative reasoning helps on complex filing questions.
Second, and more importantly, \textsc{FinSAgent} attains the best Correctness among
all multi-agent frameworks on every dataset, and does so \emph{regardless of
backbone}: under Qwen3-Max it is best or second-best on nearly all dimensions and
datasets, and its advantage is largest on the evidence-critical Factual
Consistency dimension. The gains persist across DeepSeek-v3.1-Terminus,
Kimi-K2.5, and Qwen3-Max, indicating they are attributable to the retrieval-planning
design rather than a particular generator. The error analysis of
\S\ref{sec:error-analysis} further shows \textsc{FinSAgent} suppresses the severe
failure modes that dominate baseline errors. Moreover, we run a matched-budget fairness study (Appendix~\ref{app:fairness}) in which MoA and FinDebate receive \textsc{FinSAgent}'s own per-role retrieval budget with our two mechanisms disabled. \textsc{FinSAgent} still wins Factual Consistency on all five benchmarks and Correctness on four of five, so the gains cannot be attributed to a larger number of retrieval perspectives.

\subsection{Retrieval Coverage and Component Ablation}
\label{sec:retrieval-results}

\begin{table}[t]
\centering
\small
\caption{Retrieval ablation (Macro-Recall, \%). Each row cumulatively adds a
component to the single-agent baseline; \textsc{FinSAgent} is the full system.
The final context is fixed to $\sim$35 chunks across all rows to isolate recall
\emph{quality} from quantity. \textbf{Best} bold, \underline{second-best}
underlined.}
\label{tab:retrieval_ablation}
\setlength{\tabcolsep}{4pt}
\renewcommand{\arraystretch}{1.1}
\begin{tabular}{@{}lccccc@{}}
\toprule
\textbf{Method} & \textbf{Zeekr} & \textbf{Lotus} & \textbf{FinDER} & \textbf{FinBench} & \textbf{SECQUE} \\
\midrule
Single Agent            & 28 & 41 & 50 & 58 & 60 \\
\midrule
Multi-Agent             & 31 & 35 & 57 & 71 & 74 \\
+ Feat.-gated Rerank    & 35 & 41 & \underline{58} & 71 & \underline{77} \\
+ DB-Aware Decomp.      & \underline{38} & \underline{43} & 57 & \textbf{80} & \underline{77} \\
\midrule
\textsc{FinSAgent} (Full) & \textbf{40} & \textbf{50} & \textbf{60} & \underline{76} & \textbf{83} \\
\bottomrule
\end{tabular}
\end{table}

We isolate retrieval quality by holding the final context to $\sim$35 chunks
across all configurations, so improvements reflect \emph{smarter routing}, not
more evidence. Table~\ref{tab:retrieval_ablation} shows the components are highly
complementary: multi-role retrieval drives large gains on most corpora,
feature-gated reranking filters unreliable candidates (notably lifting the hard
\texttt{Lotus} set), and database-aware decomposition resolves compositional
queries (peaking \texttt{FinanceBench} at 80). The full system achieves the best
overall coverage. A per-agent gating-strength ablation and a global SHAP
attribution (Appendix~\ref{app:gating}) further show that the gate's gains are
robust to the penalty strength $\lambda_a$ and are driven by non-semantic
features that the semantic reranker does not model.

\begin{figure}[t]
    \centering
    \includegraphics[width=\linewidth]{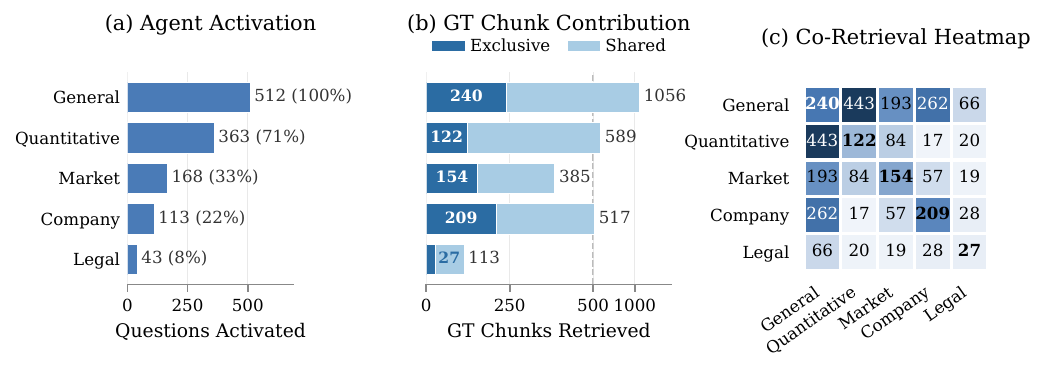}
    \caption{Agent activation and contribution. Every specialized role, including
    the low-activation Legal role, contributes exclusive ground-truth evidence.}
    \label{fig:combined_plots}
\end{figure}

The agent-activation analysis (Figure~\ref{fig:combined_plots}) validates every
role: all agents contribute exclusive ground-truth chunks to the final pool. Even
the lowest-activation \emph{Legal} agent (8\% activation) consistently retrieves
exclusive ground-truth chunks when triggered, and the co-retrieval heatmap
confirms that neighboring roles cannot absorb them---so dropping any role lowers
recall.

\paragraph{On the FinanceBench full-system dip.}
The full system dips on \texttt{FinanceBench} (80 to 76). Under our fixed
$\sim$35-chunk budget, components \emph{reallocate} rather than enlarge the budget,
so where added breadth helps little it can displace already-sufficient chunks.
This is exactly the case for \texttt{FinanceBench}: the open-source version is
corpus-shallow per entity ($\approx$41 companies, $\approx$8.5 filings each,
$\approx$73\% 10-Ks), so evidence is concentrated in a single 10-K, leaving little
complementary material for multi-perspective exploration to recover while the added
breadth competes for the same chunks. The swing is small (a handful of questions,
within run-to-run variance) and does not propagate downstream: in Table~\ref{tab:end2end},
\textsc{FinSAgent} still attains the best FinanceBench Correctness (3.59 vs.\ 3.35
next best). On corpora with a near-complete per-company filing set, role-specialized
broad retrieval pays off and every component contributes positively.

% \subsection{Are the Two Mechanisms Redundant?}
% \label{sec:query-vs-evidence}

% A natural objection is that database-aware decomposition and feature-gated
% reranking are two versions of the same ``verifier.'' They are not: they sit on
% different sides of the pipeline and help on different corpora for different
% reasons, as the component rows of Table~\ref{tab:retrieval_ablation} already
% reveal. Reading those rows on top of the Multi-Agent baseline exposes a clean
% \emph{double dissociation}. The feature gate (\emph{+Feat.-gated Rerank}) lifts
% Macro-Recall on the high-redundancy, boilerplate-heavy corpora where confident
% false positives dominate---Zeekr ($31\!\to\!35$), Lotus ($35\!\to\!41$), and
% SECQUE ($74\!\to\!77$)---but gives \emph{no} gain on \texttt{FinanceBench}
% ($71\!\to\!71$). Conversely, database-aware decomposition (\emph{+DB-Aware
% Decomp.}) drives the large \texttt{FinanceBench} jump ($71\!\to\!80$) that the
% gate cannot produce, precisely because there the binding problem is that the
% evidence was never retrieved---a gain \emph{no post-hoc verifier could produce}.
% The two mechanisms are thus not one redundant check: the query-side mechanism
% recovers missing evidence while the evidence-side gate suppresses invalid
% evidence, and the corpus structure (redundancy vs.\ shallowness) predicts which
% dominates. This is exactly the signature of two responses to a single cause
% (prior--corpus misalignment) rather than two verifiers.

\subsection{Comparison to Full-Document Long-Context Prompting}
\label{sec:longcontext}

A reasonable alternative to planned retrieval is to hand the entire document to a
frontier long-context model. We test this by giving a strong long-context model
the full 10-K with no retrieval and letting it answer directly, under the same
blind Likert protocol, on \texttt{FinanceBench}. We choose \texttt{FinanceBench}
precisely because it is the single-filing benchmark most favorable to long-context
prompting: the open-source version is shallow per entity and 10-K-dominated
($\approx$41 companies, $\approx$8.5 filings each, of which $\approx$73\% are
10-Ks), so a question's evidence is typically concentrated within one annual
report that fits comfortably in a long-context window. This document-count
distribution makes \texttt{FinanceBench} the fairest possible testbed for the
baseline; our multi-entity corpora (Lotus, Zeekr), by contrast, cannot be handled
by single-document prompting at all. The result is the LC row in
Table~\ref{tab:end2end} (FinanceBench block).

Even on its most favorable benchmark, full-document long-context prompting reaches
only 2.98 Correctness, with a strict-correct pass rate of 44.8\% (64 of 143
questions judged fully correct)---below both multi-agent baselines (MoA 3.05,
FinDebate 3.17) and 0.34 below \textsc{FinSAgent} (3.32). \textsc{FinSAgent} also
leads it on Factual Consistency (3.52 vs.\ 3.37) and Analytical Depth (3.71 vs.\
3.42); long-context only wins on surface Clarity. This supports our thesis that
grounded filing QA is a retrieval-planning problem: for this task, planned
retrieval outperforms dumping the full document into a long-context
model---before accounting for the much higher per-query token cost, and noting
that multi-entity benchmarks cannot be handled by single-document prompting at
all.

\subsection{Evaluation Validity: Human Study and Noise Sensitivity}
\label{sec:validity}

\paragraph{Blind human validation.}
To anchor the automated judge to expert judgment, we collected human annotations
on 150 answers (50 each from \textsc{FinSAgent}, MoA, and FinDebate), rated blind
and in randomized order by finance-background annotators on SECQUE's native 0/1/2
correctness scale (Table~\ref{tab:validity}). The human ranking reproduces the
automated one: \textsc{FinSAgent} 1.48 $>$ FinDebate 1.22 $>$ MoA 0.98.
\textsc{FinSAgent} is rated fully correct on 66\% of answers (vs.\ 52\%/40\%),
scores zero on 18\% (vs.\ 30\%/42\%), and has the lowest variance, so its
advantage is not an artifact of automated judging. Human and LLM-judge Correctness
scores agree strongly, with Spearman's $\rho = 0.70$ and 92.7\% within-one-point
agreement, addressing rater-reliability and judge-validity concerns.

\begin{table}[t]
\centering
\small
\caption{Blind human validation on 150 answers (SECQUE 0/1/2 scale). Human ratings
reproduce the automated ranking; human--LLM Correctness agreement is $\rho=0.70$
with 92.7\% within-one-point agreement.}
\label{tab:validity}
\setlength{\tabcolsep}{5pt}
\renewcommand{\arraystretch}{1.1}
\begin{tabular}{@{}lcccc@{}}
\toprule
\textbf{Method} & \textbf{Human mean} & \textbf{Fully correct} & \textbf{Zero} & \textbf{Std} \\
\midrule
\textsc{FinSAgent} & \textbf{1.48} & \textbf{66\%} & \textbf{18\%} & \textbf{0.79} \\
FinDebate          & 1.22 & 52\% & 30\% & 0.89 \\
MoA                & 0.98 & 40\% & 42\% & 0.91 \\
\bottomrule
\end{tabular}
\end{table}

\paragraph{RAGAS noise sensitivity.}
Noise sensitivity measures how often retrieved noise causes incorrect claims
(lower is better), directly testing whether our two mechanisms reduce irrelevant
context. To attribute the effect to the mechanisms rather than the retrieval pool,
we hand the baselines \textsc{FinSAgent}'s per-role retrieval.
Table~\ref{tab:noise} shows \textsc{FinSAgent} achieves the lowest overall noise
sensitivity (0.386) and ranks best on three of four benchmarks. Since baselines
share \textsc{FinSAgent}'s retrieval, the gap is attributable to the feature gate
and database-aware decomposition. (SECQUE is excluded as the metric requires a
gold reference answer, which it lacks.)

\begin{table}[t]
\centering
\small
\caption{RAGAS noise sensitivity (lower is better) with baselines handed
\textsc{FinSAgent}'s per-role retrieval. \textbf{Best} in bold.}
\label{tab:noise}
\setlength{\tabcolsep}{5pt}
\renewcommand{\arraystretch}{1.1}
\begin{tabular}{@{}lccccc@{}}
\toprule
\textbf{System} & \textbf{Lotus} & \textbf{Zeekr} & \textbf{FinBench} & \textbf{FinDER} & \textbf{Mean} \\
\midrule
\textsc{FinSAgent} & \textbf{0.357} & \textbf{0.347} & \textbf{0.307} & 0.532 & \textbf{0.386} \\
MoA                & 0.388 & 0.470 & 0.392 & 0.591 & 0.460 \\
FinDebate          & 0.358 & 0.452 & 0.365 & \textbf{0.494} & 0.417 \\
\bottomrule
\end{tabular}
\end{table}

\subsection{System Overhead and Latency}
\label{sec:overhead}

\begin{figure}[t]
    \centering
    \includegraphics[width=\linewidth]{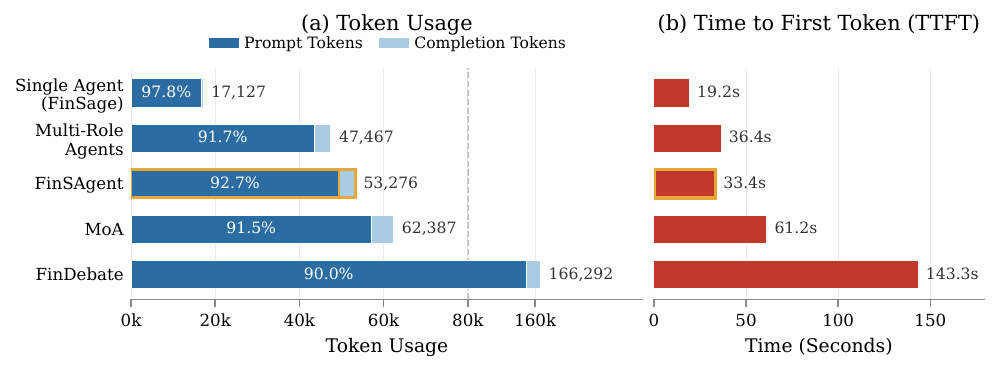}
    \caption{System overhead. (a) Total token usage split into prompt/completion
    tokens (inner percentages are the prompt-token ratio). (b) Time-to-first-token
    (TTFT). \textsc{FinSAgent} achieves far lower latency than external multi-agent
    configurations while keeping token consumption efficient.}
    \label{fig:system_overhead}
\end{figure}

Figure~\ref{fig:system_overhead} summarizes token usage and time-to-first-token
(TTFT). Although database-aware decomposition raises average token usage to
64{,}914, TTFT actually drops to 50.10\,s (from 52.14\,s for the plain multi-role
setup), because feature-gated reranking tightly bounds the orchestrator's context
during synthesis. \textsc{FinSAgent} delivers specialized reasoning at roughly
39\% of FinDebate's token cost and 35\% of its latency.

\begin{table}[t]
\centering
\small
\caption{Per-phase latency and token breakdown for \textsc{FinSAgent}. Parallel
sub-agent phases are reported as their union (max over concurrent calls). Totals
for the baselines are shown for reference.}
\label{tab:latency}
\setlength{\tabcolsep}{5pt}
\renewcommand{\arraystretch}{1.1}
\begin{tabular}{@{}lrr@{}}
\toprule
\textbf{Phase} & \textbf{Latency (s)} & \textbf{Tokens} \\
\midrule
Agent activation      & 1.91  & 934 \\
Query decompose       & 4.39  & 18{,}332 \\
Agent sub-answer      & 23.54 & 33{,}516 \\
Final synthesize      & 5.84  & 495 \\
Retrieval + rerank    & 2.33  & -- \\
\midrule
\textbf{\textsc{FinSAgent} total} & \textbf{38.01} & \textbf{53{,}276} \\
\midrule
MoA total             & 71.42  & 62{,}387 \\
FinDebate total       & 155.44 & 166{,}292 \\
\bottomrule
\end{tabular}
\end{table}

The per-phase breakdown (Table~\ref{tab:latency}) explains the efficiency: the
planning and retrieval overhead is small, and roughly 62\% of total latency
(23.54\,s) sits in a single \emph{parallel} sub-answer pass. The baselines instead
spend their time in sequential LLM rounds---MoA runs two sequential answer rounds
plus synthesis, and FinDebate runs four sequential debate stages plus
synthesis---so \textsc{FinSAgent} runs at roughly 53\% of MoA's latency and 24\%
of FinDebate's, at about one-third of FinDebate's token cost. It achieves role
specialization through parallelism rather than sequential refinement.

\subsection{Case Study: Correcting a Boilerplate False Positive}
\label{sec:case-study}

We illustrate the mechanism on a concrete case (Zeekr run,
idx=59). For ``\emph{What are Zeekr's main risks?}'', the ground truth centers on
company-specific operating risks: European expansion, China sales concentration,
premium-BEV competition, supplier dependence, and product execution. The MoA
baseline retrieves risk-factor paragraphs but its semantic reranker ranks generic
regulatory boilerplate (PRC overseas-listing rules, data/security regulations,
Cayman holding-company structure) above the company-specific evidence, so the
synthesized answer over-weights issuer-level compliance language and under-weights
the risks that actually drive the business.

\textsc{FinSAgent}'s feature-gated reranker applies
$\tilde{s} = s^{\mathrm{rr}}\cdot(1 - \lambda\,\hat{r})$ with $\lambda=0.30$ and
$\hat{r}=1-\hat{p}$. Boilerplate chunks receive high risk ($\hat{r}\in[0.89,0.98]$)
and are demoted, while company-specific ground-truth chunks (China concentration,
Europe expansion) receive lower risk ($\hat{r}\in[0.75,0.78]$) and enter the top
slots. Semantic similarity alone cannot distinguish boilerplate from
company-specific evidence when both are about ``risk''; the learned gate carries
the signal the reranker misses. In aggregate this is net-positive: the gate more
often \emph{reorders} evidence than changes coverage, and the same false-positive
suppression drives the error reductions in \S\ref{sec:error-analysis}.

\subsection{Error Analysis}
\label{sec:error-analysis}

Using a fine-grained error taxonomy (Appendix~\ref{app:error_distribution}),
\textsc{FinSAgent} achieves the lowest total error count on both \texttt{SECQUE}
and \texttt{ZEEKR}. Two reductions are directly tied to our two mechanisms. Query
misunderstanding (D1), up to 31\% of baseline errors, drops to 15\% (SECQUE) and
6\% (ZEEKR), reflecting database-aware decomposition. On multi-entity
\texttt{ZEEKR}, evidence-fusion failure (B4) falls from 23--25\% in retrieval-only
baselines to 15\%. Numerical-precision (C1) errors appear only in \textsc{FinSAgent}; baselines fail earlier in the pipeline and never reach the point where precision matters.
We read this as evidence that the remaining bottleneck has moved downstream,
from retrieval and comprehension to numeric fidelity.
\section{Conclusion}
\label{sec:conclusion}

We presented \textsc{FinSAgent}, a multi-agent framework for evidence-grounded
question answering over SEC filings. Our central claim is conceptual rather than
integrative: SEC filing QA is best understood as corpus-aligned retrieval
planning, and the binding constraint is \emph{prior--corpus misalignment}, which is a
single cause that surfaces symmetrically at query formation and evidence
selection. \textsc{FinSAgent} corrects both ends with one principle, injecting
corpus-side conditioning through database-aware query decomposition before
retrieval and a learned feature gate that separates evidential validity from
semantic similarity after retrieval, on top of role-specialized agents anchored to
the mandated 10-K item structure. Across five benchmarks, controlled ablations, a
matched-budget fairness study, a blind human validation, a full-document
long-context comparison, and RAGAS noise sensitivity, \textsc{FinSAgent} improves
retrieval coverage and answer correctness, and a mechanistic SHAP analysis
confirms that non-semantic features carry the validity signal the reranker misses.
A double-dissociation ablation shows the two corrective mechanisms help on
different corpora for different reasons, evidence that they are two responses to
one cause rather than a redundant verifier.

\clearpage

\section*{Limitations and future work}
Our evidence for the two mechanisms currently rests on full-system ablations and
fixed-pool comparisons. To further isolate the design motivation, we plan to add
direct \emph{separated} baselines: on the query side, comparing database-aware
decomposition against a lightweight terminology-adaptation agent and a static
corpus-overview prompt under matched cost; and on the evidence side, comparing the
feature-gated reranker against an explicit claim/NLI-style passage verifier that
converts sub-queries or draft answers into atomic hypotheses and reranks by
entailment, inspired by FEVER-style claim--evidence
verification~\citep{thorne2018fever} and SummaC-style sentence-level NLI
aggregation~\citep{laban2022summac}. These would test whether the gains stem from
non-semantic evidence-validity signals beyond semantic or entailment-based
verification. Fine-grained numerical precision (the residual C1 errors of
\S\ref{sec:error-analysis}) remains the primary open problem, and extending the
version-aware treatment of supersession into an explicit temporal reasoning module
is a natural next step. More broadly, the prior--corpus misalignment framing and
the feature-gating mechanism should transfer to any corpus that is standardized
and redundancy-heavy, which we leave to future study.

\begin{acks}
This work was carried out in collaboration with CG Matrix Technology Limited
and SimpleWay.AI. We are grateful to both organizations for providing
computational resources, datasets, and domain expert guidance that made this
work possible.
\end{acks}

\section*{Ethics and Privacy Statement}
This research introduces \textsc{FinSAgent}, a multi-agent system for financial
question answering over public SEC filings. Because financial analysis involves
high-stakes decision-making, we emphasize that \textsc{FinSAgent} is an analytical
research tool; its outputs do not constitute professional financial, investment,
or legal advice. To mitigate the ethical risks of model hallucination in the
financial domain, the framework is explicitly designed to improve evidence
reliability and factual grounding rather than fluency alone. All experiments use
publicly available corporate disclosures and proprietary datasets released to us
for research use; no non-public or personally identifiable information is
processed. We view the improved false-positive suppression and interpretable
gating as a net benefit for auditability, while noting that any deployment in
real financial workflows must retain a human expert in the loop.

\section*{Use of Large Language Models}
We used a general-purpose assistant to check grammar and refine phrasing. All
technical claims, experiments, and analyses were designed and verified by the
authors. We additionally use DeepSeek-v3.1-Terminus, Kimi-K2.5, and Qwen3-Max as
generation backbones and answer-quality judges in our experiments, as detailed in
Section~\ref{sec:experiments}.

\bibliographystyle{ACM-Reference-Format}
\bibliography{finsagent}

\appendix
\section{Datasets, Preprocessing, and Baselines}
\label{app:preprocess}

\subsection{Corpus Construction}
Each filing is parsed with \texttt{MinerU}~\citep{wang2024mineru} into 200-word
chunks tagged by modality. The pipeline proceeds as follows.
\begin{enumerate}[leftmargin=1.5em]
    \item \textbf{PDF parsing.} Documents are parsed into structured JSON
    representations with \texttt{MinerU}.
    \item \textbf{Formatting and deduplication.} Parsed text is converted to a
    unified chunk format with page and section metadata; empty entries are
    filtered and near-duplicate passages removed (BGE-M3 cosine similarity above a
    threshold $\tau_{\mathrm{sim}}$). Non-textual elements (tables, figures) are
    isolated for specialized processing.
    \item \textbf{Contextual enrichment.} An LLM resolves anaphoric references
    within the local section context (co-reference resolution, so chunks are
    self-contained), and a section-level summary is attached to every chunk.
    \item \textbf{Re-chunking.} Text is divided into retrieval-friendly units of
    uniform length, preserving sentence boundaries; each chunk receives a
    normalized identifier.
    \item \textbf{Multimodal processing.} Figures are converted to structured
    caption text (via a compact multimodal model) and thereafter treated as text
    chunks. Tables use a dedicated branch: a strong model generates a summary
    (title + data-trend description) for retrieval while the full table is kept as
    HTML in metadata; at inference the summary is retrieved and its HTML pulled
    into context.
    \item \textbf{Index construction.} Chunks are indexed in parallel with BGE-M3
    into dense vectors in Chroma, plus a sparse inverted BM25 index and separate
    indices for section summaries and reconstructed tables, supporting the
    multi-path (sparse / dense / summary / table) retrieval the agents use.
\end{enumerate}

\subsection{SECQUE Sampling}
For \texttt{SECQUE} we use stratified sampling that preserves the benchmark's
native question-type distribution rather than uniform random sampling. From the
565-question pool across four question types, we draw 100 proportionally, with per-type
counts of 39, 33, 15, and 13 (Comparison \& Trend Analysis, Ratio Analysis, and two
risk/insight types). This keeps the evaluation subset representative of SECQUE's
type mix.

\subsection{Baseline Setup}
\label{app:baselines}

\paragraph{Naive RAG.}
A non-agentic retrieve-then-generate baseline. Given a question, it retrieves a
small set of chunks from a mixed collection and performs a single LLM generation
step conditioned only on the retrieved evidence. We evaluate a dense-only variant
(top-$k$ from a FAISS dense retriever) and a 2-path variant (dense $+$ BM25 after
document-level deduplication). The model is instructed to answer solely from the
retrieved context and to abstain when evidence is insufficient. No agent routing,
multi-round interaction, tool calling, or debate is used.

\paragraph{FinGPT.}
A single-agent, tool-augmented system built on an agent SDK with Model Context
Protocol (MCP) integration. It classifies the question by lexical pattern matching
and routes it through an MCP-first path (numerical queries), an iterative-research
path (complex queries decomposed into typed sub-questions with gap detection), or a
single-web-search path (simple/qualitative queries). For a fair document-grounded
comparison, we restrict tool access to leave SEC-EDGAR as the only active
structured source, disabling market-data and filesystem MCP servers.

\paragraph{FinDebate.}
A shared-retrieval, multi-role debate baseline. Non-English questions are
translated into concise English retrieval queries, and a single multi-path
retrieval step builds a shared evidence context (as in FinSage) provided to all
downstream agents; no specialist retrieves further. Five specialist roles (earnings
analyst, market predictor, risk analyst, sentiment analyst, valuation analyst) each
produce an initial report, then undergo a three-stage \emph{Trust}/\emph{Skeptic}/\emph{Leader}
debate, after which a synthesizer aggregates the five leader reports.

\paragraph{Mixture of Agents (MoA).}
A two-round mixture-of-agents baseline with shared retrieval. As in FinDebate, a
single retrieval step yields a shared context. Four specialists (market, company,
quantitative, legal) plus the general agent each produce a first-round answer; all
first-round outputs are then exposed to every specialist, which revises its answer
using peer outputs only when supported by the shared evidence. A final synthesis
module combines the second-round drafts.

\section{Matched-Budget Fairness Study}
\label{app:fairness}

This appendix details the matched-budget fairness study summarized in
\S\ref{sec:e2e}. To rule out that \textsc{FinSAgent}'s gains merely reflect having
multiple retrieval perspectives, we re-run MoA and FinDebate under
\textbf{matched per-role retrieval}: each baseline specialist retrieves at the same
per-role budget as its role-matched \textsc{FinSAgent} counterpart (per-agent
retrieve\_top\_k$=10$, rerank\_top\_k$=5$), with database-aware decomposition and
feature-gated reranking turned off, and a baseline agent skipped when the mapped
\textsc{FinSAgent} role is inactive (mirroring dynamic activation). All systems
share the same backbone. Table~\ref{tab:fairness} counts per-dimension first-place
wins across the five benchmarks.

\begin{table}[h]
\centering
\small
\caption{First-place wins across five benchmarks under matched per-role retrieval
(ties counted for all tied methods). Even when handed \textsc{FinSAgent}'s own
retrieval budget, baselines cannot match it on the evidence-critical dimensions.}
\label{tab:fairness}
\setlength{\tabcolsep}{5pt}
\renewcommand{\arraystretch}{1.1}
\begin{tabular}{@{}lccc@{}}
\toprule
\textbf{Dimension} & \textbf{\textsc{FinSAgent}} & \textbf{FinDebate} & \textbf{MoA} \\
\midrule
Information Coverage   & 2          & 0            & 3 \\
Reasoning Chain        & 0          & 3            & 2 \\
Factual Consistency    & \textbf{5} & 0            & 0 \\
Clarity of Expression  & 1          & 4            & 0 \\
Analytical Depth       & 0          & 3\,(+1 tie)  & 1\,(+1 tie) \\
Correctness            & \textbf{4} & 0            & 1 \\
\midrule
\textbf{Total (incl.\ ties)} & \textbf{12} & 11\,(+1 tie) & 7\,(+1 tie) \\
\bottomrule
\end{tabular}
\end{table}

At a matched retrieval budget, \textsc{FinSAgent} wins Factual Consistency on all
five benchmarks and Correctness on four of five. The margins over the best
baseline on these two evidence-critical dimensions are consistently positive:
Factual Consistency improves by $+0.02$ to $+0.40$ across the five benchmarks, and
Correctness by up to $+0.48$, remaining positive on four of five (with a single
$-0.08$ on Lotus). The baselines become competitive on breadth and presentation
dimensions (coverage, clarity, depth), but this does not carry over to grounding
or correctness. In other words, extra deliberation cannot offset retrieval
deficiencies: the advantage comes from database-aware decomposition and
feature-gated reranking, not from having more retrieval perspectives.

\section{Gating Model}
\label{app:gating}

This appendix details the feature specification, training, and interpretation of
the feature gate used in feature-gated reranking (\S\ref{sec:retrieval}).

\subsection{Feature Specification}
\label{app:gating_features}

The gate operates on a feature vector $\mathbf{x}(c, q_{a,j})$ built for each
candidate chunk $c$ and sub-query $q_{a,j}$. We use 31 features (including the
reranker score) in five groups (Table~\ref{tab:gating_features}). \emph{Retrieval-path}
features record provenance and raw scores from each path (sparse, dense, summary,
table) and within-path rank positions. \emph{Lexical-overlap} features capture
token-, number-, and year-overlap statistics between query and chunk.
\emph{Chunk-metadata} features encode content type, document position, and
publication year. \emph{Query-context} features provide query-level signals such
as language. The \emph{reranker} feature is the cross-encoder score itself, which
lets the gate learn non-linear interactions between semantic similarity and the
retrieval-side signals.

\begin{table}[h]
\centering
\small
\caption{Feature groups used by the gate. Categorical features are marked
$^\dagger$.}
\label{tab:gating_features}
\begin{tabularx}{\linewidth}{@{}l c >{\raggedright\arraybackslash}X@{}}
\toprule
\textbf{Group} & \textbf{\#} & \textbf{Features} \\
\midrule
Retrieval path & 15 &
\texttt{retrieval\_path}$^\dagger$, \texttt{num\_retrieval\_paths},
\texttt{has\_\{faiss,bm25,title\_summary,table\}},
\texttt{\{faiss,bm25,title\_summary,table\}\_score}, \texttt{\{faiss,bm25,title\_summary,table\}\_rank}, \texttt{min\_rank} \\
\midrule
Lexical overlap & 11 &
\texttt{\{query,chunk,title\_summary\}\_token\_len},
\texttt{token\_overlap\_\allowbreak\{count,\allowbreak ratio\_query,\allowbreak ratio\_chunk\}}, \texttt{token\_jaccard},
\texttt{\{query,chunk\}\_number\_count}, \texttt{number\_overlap\_count},
\texttt{year\_overlap\_count} \\
\midrule
Chunk metadata & 3 & \texttt{chunk\_type}$^\dagger$, \texttt{page\_number}, \texttt{doc\_year} \\
\midrule
Query context & 1 & \texttt{query\_language}$^\dagger$ \\
\midrule
Reranker & 1 & \texttt{decoder\_score} \\
\bottomrule
\end{tabularx}
\end{table}

\subsection{Training}
\label{app:gating_training}
We train a LightGBM binary classifier (binary cross-entropy) on labeled
chunk--query pairs. To prevent data leakage, the training set is built exclusively
via an automated in-distribution pipeline, never from the evaluation sets. Using
targeted hard-negative mining, an LLM synthesizes queries from ground-truth
narrative and table chunks; these are processed through the full retrieval and
reranking pipeline; within the top-$k$ results, ground-truth matches are labeled
positive ($y=1$) and the remaining retrieved chunks serve as competitive hard
negatives, downsampled to an approximate 20\% positive rate. Categorical features
are handled natively by LightGBM's optimal-split algorithm. The predicted risk
score used by the gate is $\hat{r} = 1 - \hat{p}_{\mathrm{relevant}}$.

\subsection{Gating Strength Ablation}
\label{sec:gating-ablation}

\begin{table}[h]
\centering
\small
\caption{Macro-Recall (\%) on Lotus across gating strengths $\lambda_a$. $\Delta$
is the gain of the best $\lambda_a>0$ over the ungated baseline ($\lambda_a=0$).}
\label{tab:lambda}
\setlength{\tabcolsep}{6pt}
\renewcommand{\arraystretch}{1.1}
\begin{tabular}{@{}lcccc@{}}
\toprule
$\boldsymbol{\lambda_a}$ & \textbf{Company} & \textbf{Legal} & \textbf{Market} & \textbf{Quant} \\
\midrule
0.0 & 17.4 & 50.9 & 22.7 & 32.6 \\
0.1 & 23.2 & 54.4 & 23.0 & 36.3 \\
0.2 & 23.2 & \textbf{55.0} & \textbf{23.0} & 36.1 \\
0.3 & 23.0 & 55.0 & 22.9 & 35.6 \\
0.6 & \textbf{23.5} & 55.0 & 22.7 & \textbf{37.0} \\
\midrule
$\Delta$ & \textit{+6.1} & \textit{+4.1} & \textit{+0.3} & \textit{+4.4} \\
\bottomrule
\end{tabular}
\end{table}

Table~\ref{tab:lambda} ablates the per-agent gating strength $\lambda_a$ on Lotus.
Enabling the gate ($\lambda_a>0$) consistently improves retrieval over pure
reranking ($\lambda_a=0$); the Company and Quant agents benefit most, as the gate
suppresses topically related but invalid chunks. Performance is robust across
$\lambda_a$: most of the gain appears already at $\lambda_a=0.1$, with 0.2--0.6
adding little. This plateau shows a mild correction suffices to demote the most
confident false positives without aggressive tuning; in practice we set
$\lambda_a\in[0.1,0.3]$ per agent.

\subsection{Feature Importance}
\begin{figure}[h]
    \centering
    \includegraphics[width=0.85\linewidth]{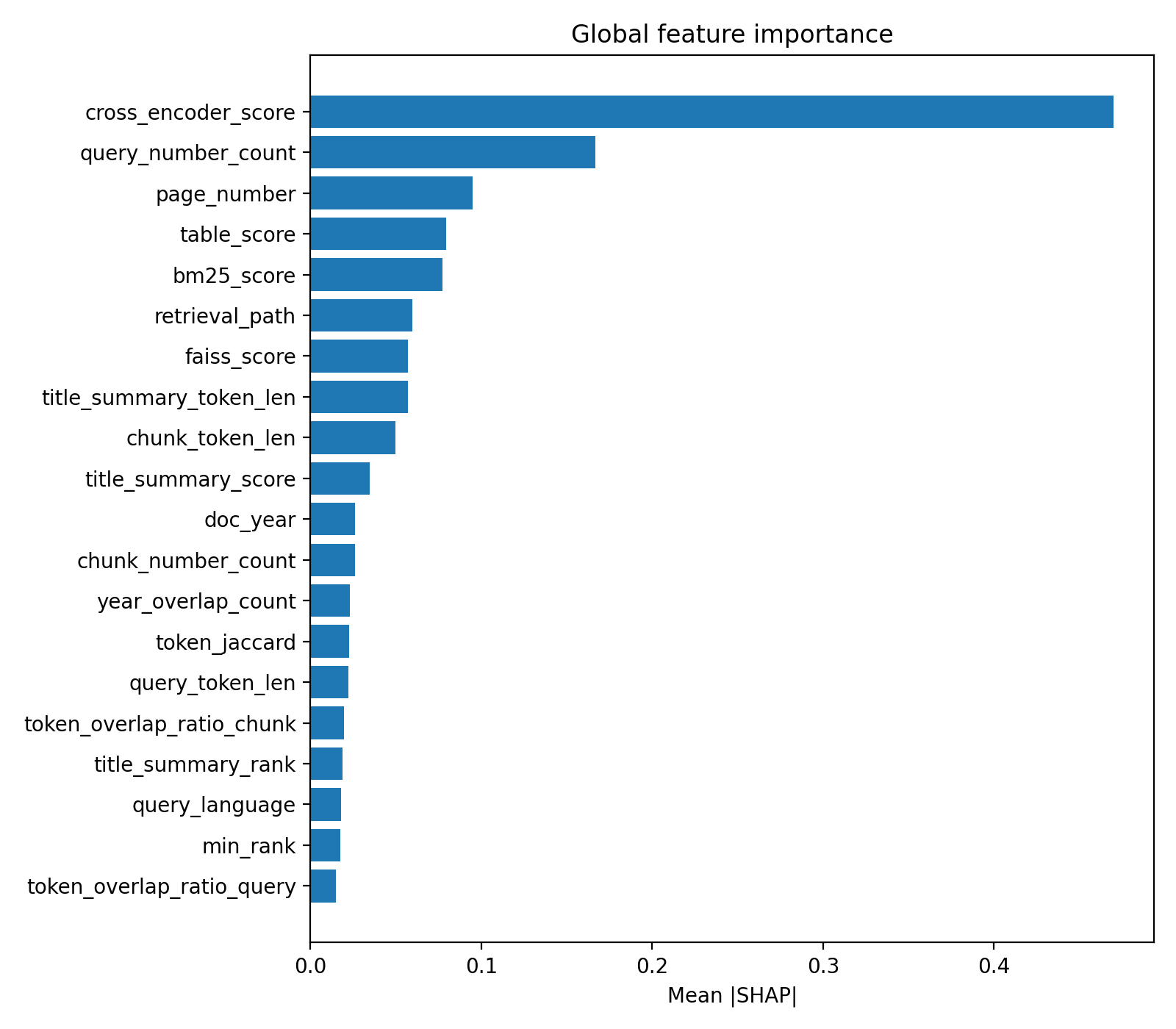}
    \caption{Global SHAP feature importance. The gate relies primarily on the
    reranker score, supplemented by secondary retrieval and lexical features that
    the cross-encoder does not model.}
    \label{fig:shap}
\end{figure}

Figure~\ref{fig:shap} shows global SHAP importance. The reranker score dominates
(mean $|\text{SHAP}|\approx 0.47$), followed by \texttt{query\_number\_count}
(0.17) and \texttt{page\_number} (0.09). Retrieval-path scores
(\texttt{table\_score}, \texttt{bm25\_score}, \texttt{faiss\_score}) and length
features form a second tier. This confirms that the gate primarily \emph{augments}
the strong semantic reranker signal with complementary, \emph{non-semantic}
retrieval and lexical cues---the mechanistic basis for the semantic--validity
decoupling that motivates the feature gate (\S\ref{sec:retrieval}), and the reason
the gating-strength ablation above yields consistent gains.

\section{Error Taxonomy and Distribution}
\label{app:error_distribution}

For fine-grained failure analysis we assign each incorrect response a primary
error subtype from one of three groups.

\paragraph{Group B: Generation-related errors.}
\begin{itemize}[leftmargin=1.4em]
    \item \textbf{B1 (Hallucination):} information contradicting or unsupported by
    retrieved evidence---fabricating numbers/categories (B1-1), inventing entity
    attributes (B1-2), unsupported comparisons (B1-3), fabricated temporal trends
    (B1-4).
    \item \textbf{B2 (Contradicts evidence):} internal logical inconsistency or
    direct conflict with previously cited evidence.
    \item \textbf{B3 (Excessive inference):} over-extrapolation beyond what the
    documents justify.
    \item \textbf{B4 (Evidence-fusion failure):} failing to synthesize
    complementary or conflicting evidence into a coherent conclusion.
\end{itemize}

\paragraph{Group C: Finance-specific numeric and semantic errors.}
\begin{itemize}[leftmargin=1.4em]
    \item \textbf{C1 (Numerical precision):} rounding errors, tolerance
    mismatches, percentages vs.\ basis points.
    \item \textbf{C2 (Units and scales):} millions vs.\ billions, currency
    mismatch, ratios vs.\ absolute values.
    \item \textbf{C3 (Time mismatch):} data from the wrong fiscal period.
    \item \textbf{C4 (Computation logic):} correct raw data but incorrect
    arithmetic or formula.
\end{itemize}

\paragraph{Group D: Query and context errors.}
\begin{itemize}[leftmargin=1.4em]
    \item \textbf{D1 (Query misunderstanding):} missing the prompt's objective,
    including intent (D1-1), entity (D1-2), and metric (D1-3) misidentification.
    \item \textbf{D2 (Context-window abuse):} losing critical information to
    context-length limits or failing to prioritize relevant evidence.
\end{itemize}

\begin{figure}[h]
    \centering
    \includegraphics[width=\linewidth]{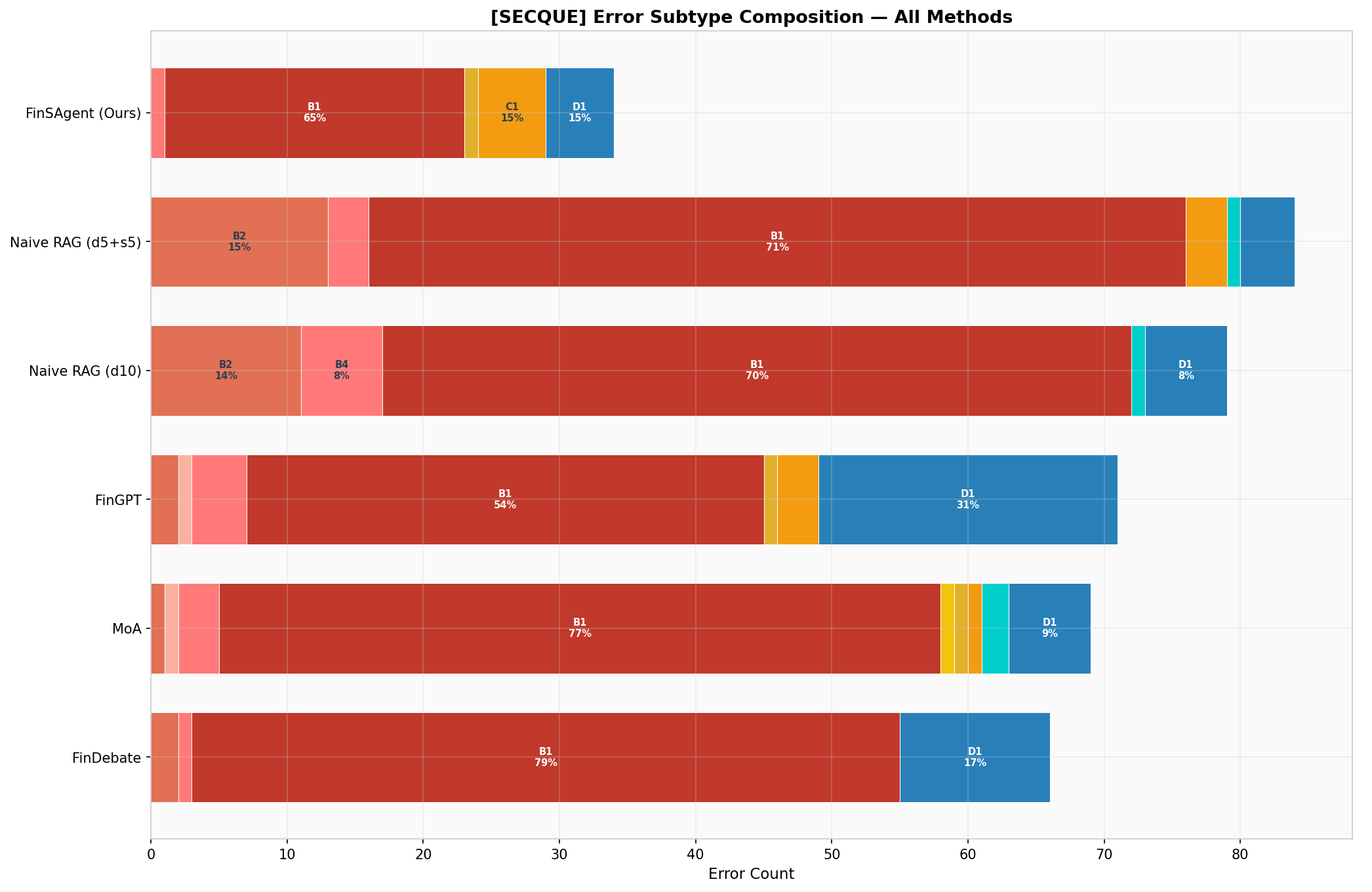}
    \caption{Error subtype composition on \texttt{SECQUE}. \textsc{FinSAgent}
    achieves the lowest error count with a qualitatively shifted profile: C1
    emerges while D1 is largely suppressed.}
    \label{fig:error_secque}
\end{figure}

\begin{figure}[h]
    \centering
    \includegraphics[width=\linewidth]{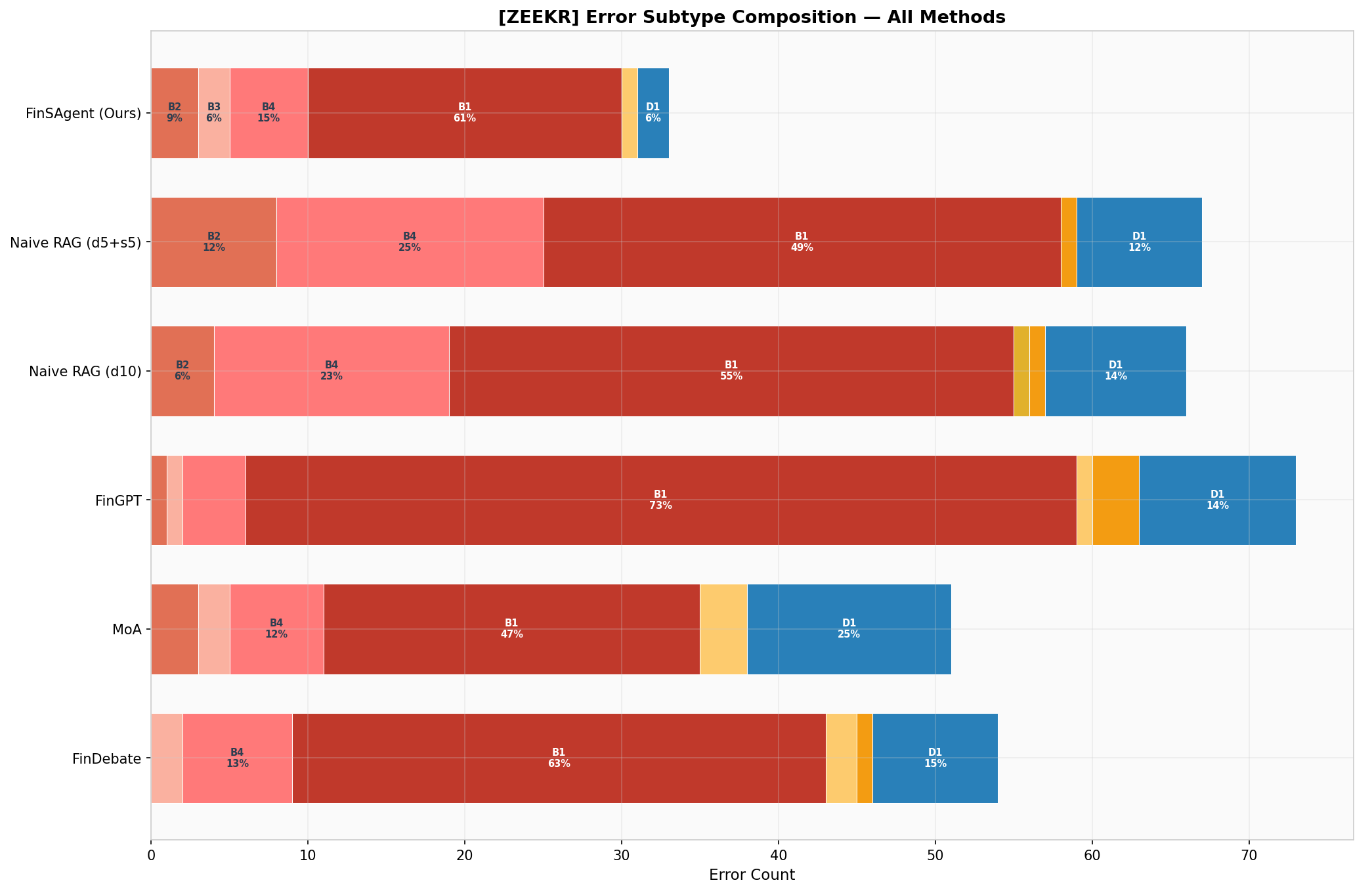}
    \caption{Error subtype composition on \texttt{ZEEKR}. The multi-entity setting
    amplifies B4 in retrieval-only baselines; \textsc{FinSAgent} reduces both B4
    and D1.}
    \label{fig:error_zeekr}
\end{figure}

\paragraph{Results.}
Applying this taxonomy to all incorrect responses on \texttt{SECQUE} and
\texttt{ZEEKR} (Figures~\ref{fig:error_secque}, \ref{fig:error_zeekr}),
\textsc{FinSAgent} attains the lowest total error count on both: roughly 26 errors
on \texttt{SECQUE} (vs.\ 68--82 for baselines) and 33 on \texttt{ZEEKR} (vs.\
55--68). Hallucination (B1) remains the dominant mode across systems, but
\textsc{FinSAgent}'s absolute B1 count is substantially lower. Query
misunderstanding (D1), up to 31\% of baseline errors (e.g., FinGPT on
\texttt{SECQUE}; MoA at 25\% on \texttt{ZEEKR}), drops to 15\% and 6\%
respectively, reflecting database-aware decomposition. On multi-entity
\texttt{ZEEKR}, evidence-fusion failure (B4) is prominent in Naive RAG variants
(23--25\%) but reduced to 15\% under \textsc{FinSAgent}. Finally, C1 (numerical
precision) errors---15\% on \texttt{SECQUE} yet absent in baselines---indicate
that \textsc{FinSAgent} resolves the upstream retrieval and comprehension failures,
leaving fine-grained numeric precision as the main remaining frontier.

\section{Optional Conversational Memory for Multi-Turn Filing QA}
\label{app:memory}

The core \textsc{FinSAgent} pipeline targets single-turn, evidence-grounded QA. To
support follow-up questions, we add an optional conversational-memory module. Let
$H_t=\{(u_1,y_1),\dots,(u_{t-1},y_{t-1})\}$ be the dialogue history up to turn $t$
and $M_t = M(H_t,u_t)$ the retrieved memory signal for the current question $u_t$.
We maintain both short-term dialogue memory and case-level long-term memory. With
memory enabled, each agent constructs $z_a^{(t)} = g_a(u_t, M_t; \pi_a)$, after
which decomposition and retrieval proceed as in the main pipeline. Memory serves
only as a contextual prior: final answers remain grounded in newly retrieved
filing evidence.

\paragraph{Architecture.}
The \emph{MemoryManager} uses a URI-abstracted storage layer for horizontal
scalability, an asynchronous queue-based dual-write mechanism for peak shaving, a
multi-granularity memory hierarchy (L0 abstract / L1 overview / L2 content) to
avoid context overflow, and a rule-based fallback for robustness. To remain
available under generative-API timeouts or schema-validation failures, the primary
LLM extractor $f_{\mathrm{LLM}}$ degrades gracefully to a heuristic extractor
$f_{\mathrm{Rule}}$:
\[
\mathcal{M}(x) = \begin{cases}
f_{\mathrm{LLM}}(x), & \text{if } f_{\mathrm{LLM}}(x)\in\mathcal{V} \wedge \Delta t < \tau_{\mathrm{timeout}}\\
f_{\mathrm{Rule}}(x), & \text{otherwise},
\end{cases}
\]
where $\mathcal{V}$ is the set of structurally valid JSON schemas and
$\tau_{\mathrm{timeout}}$ the maximum permissible latency.

\begin{figure}[h]
    \centering
    \includegraphics[width=0.72\linewidth]{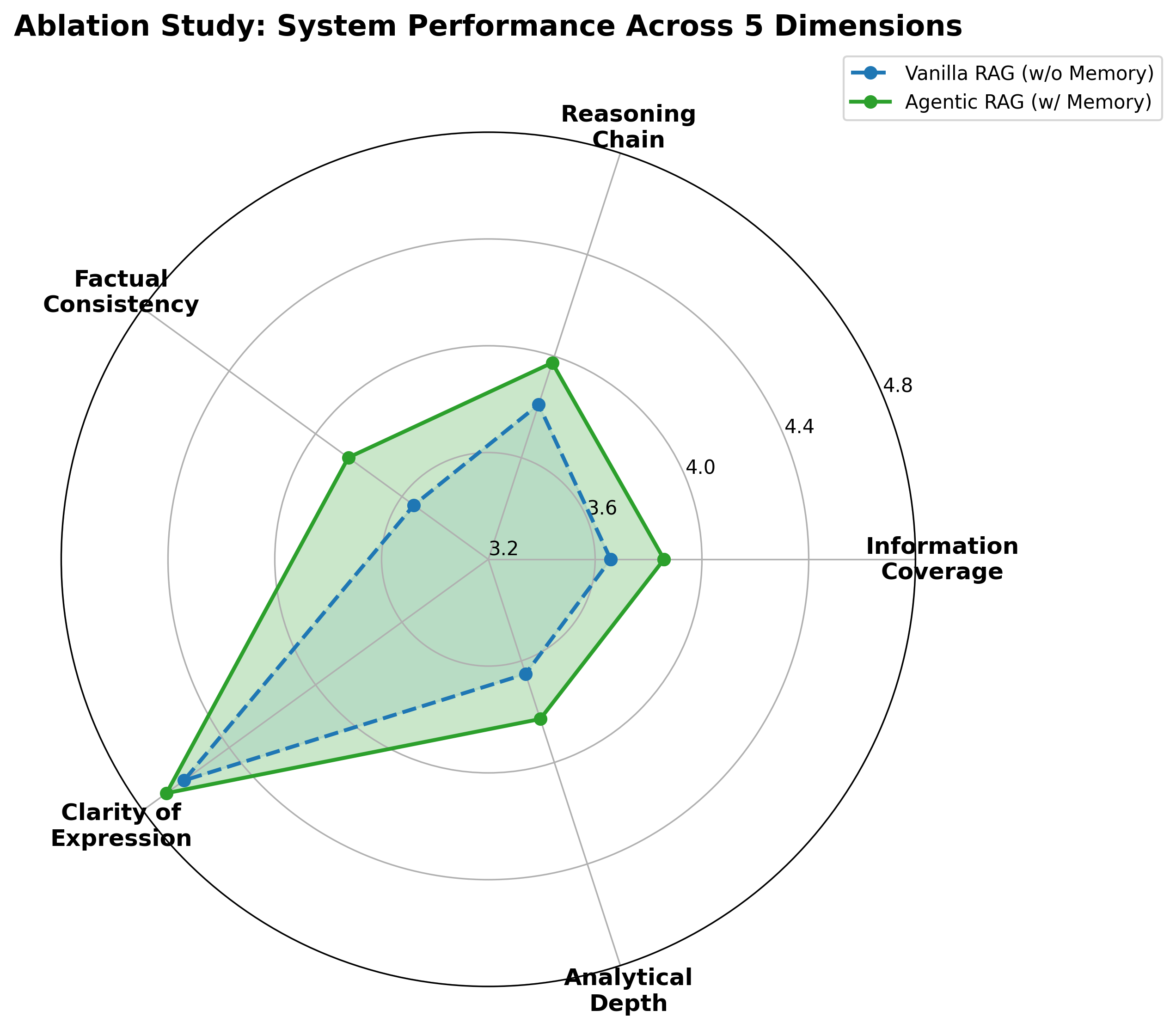}
    \caption{Multi-turn performance. The memory module improves all five measured
    dimensions.}
    \label{fig:memory_radar}
\end{figure}

\paragraph{Effect.}
An end-to-end ablation on 310 complex, multi-turn financial QA pairs sampled from
authentic user conversational logs (Figure~\ref{fig:memory_radar}) shows the module
improves all measured axes, most notably Factual Consistency ($+0.303$), and lifts
the strict-correct pass rate by 6.77\% (to 49.35\%). By injecting structured,
multi-grained history into the reasoning chain, the module acts as a semantic
anchor that suppresses hallucination in deep multi-hop QA, while the L0--L2
hierarchy preserves token budget for core reasoning and improves retrieval of
long-tail historical entities across turns.

\section{System Implementation}
\label{app:implementation}

\subsection{Multi-Path Retrieval}
\label{app:retrieval}

\textsc{FinSAgent} follows a multi-facet retrieval scheme with a sparse retriever
$R^{\mathrm{sp}}$ (BM25), a dense retriever $R^{\mathrm{de}}$ (FAISS over chunk
embeddings), a summary retriever $R^{\mathrm{sm}}$ (FAISS over section-summary
embeddings), and a table retriever $R^{\mathrm{table}}$ (FAISS over table-summary
embeddings). For each agent $a\in\mathcal{A}$ and sub-query
$q_{a,j}\in\mathcal{Q}_a$, the candidate set over the agent scope $\Omega_a$ is
\[
C_{a,j}
=
R^{\mathrm{sp}}_{k}(q_{a,j};\Omega_a)
\cup
R^{\mathrm{de}}_{k}(q_{a,j};\Omega_a)
\cup
R^{\mathrm{sm}}_{k}(q_{a,j};\Omega_a)
\cup
R^{\mathrm{table}}_{k}(q_{a,j};\Omega_a),
\]
where each path returns its top-$k$ chunks under a distinct scoring function:
\begin{align*}
    R^{\mathrm{sp}}_{k}(q_{a,j};\Omega_a) &= \operatorname*{top\text{-}}k_{c \in \Omega_a} \big( \mathrm{BM25}(q_{a,j}, c) \big), \\
    R^{\mathrm{de}}_{k}(q_{a,j};\Omega_a) &= \operatorname*{top\text{-}}k_{c \in \Omega_a} \big( \mathrm{sim}(E(q_{a,j}), E(c)) \big), \\
    R^{\mathrm{sm}}_{k}(q_{a,j};\Omega_a) &= \operatorname*{top\text{-}}k_{c \in \Omega_a} \big( \mathrm{sim}(E(q_{a,j}), E(s_{c})) \big), \\
    R^{\mathrm{table}}_{k}(q_{a,j};\Omega_a) &= \operatorname*{top\text{-}}k_{c \in \Omega_a} \big( \mathrm{sim}(E(q_{a,j}), E(t_{c})) \big),
\end{align*}
where $E(\cdot)$ is the embedding encoder, $\mathrm{sim}(\cdot,\cdot)$ is cosine
similarity, $s_c$ is the section summary of chunk $c$, and $t_c$ is its associated
table content. This preserves the complementarity of lexical, dense, and
summary-level retrieval while restricting each agent to its evidence scope.

\subsection{Multi-Agent Prompts}
\label{app:multi-agent-prompt}

\begin{figure*}[t]
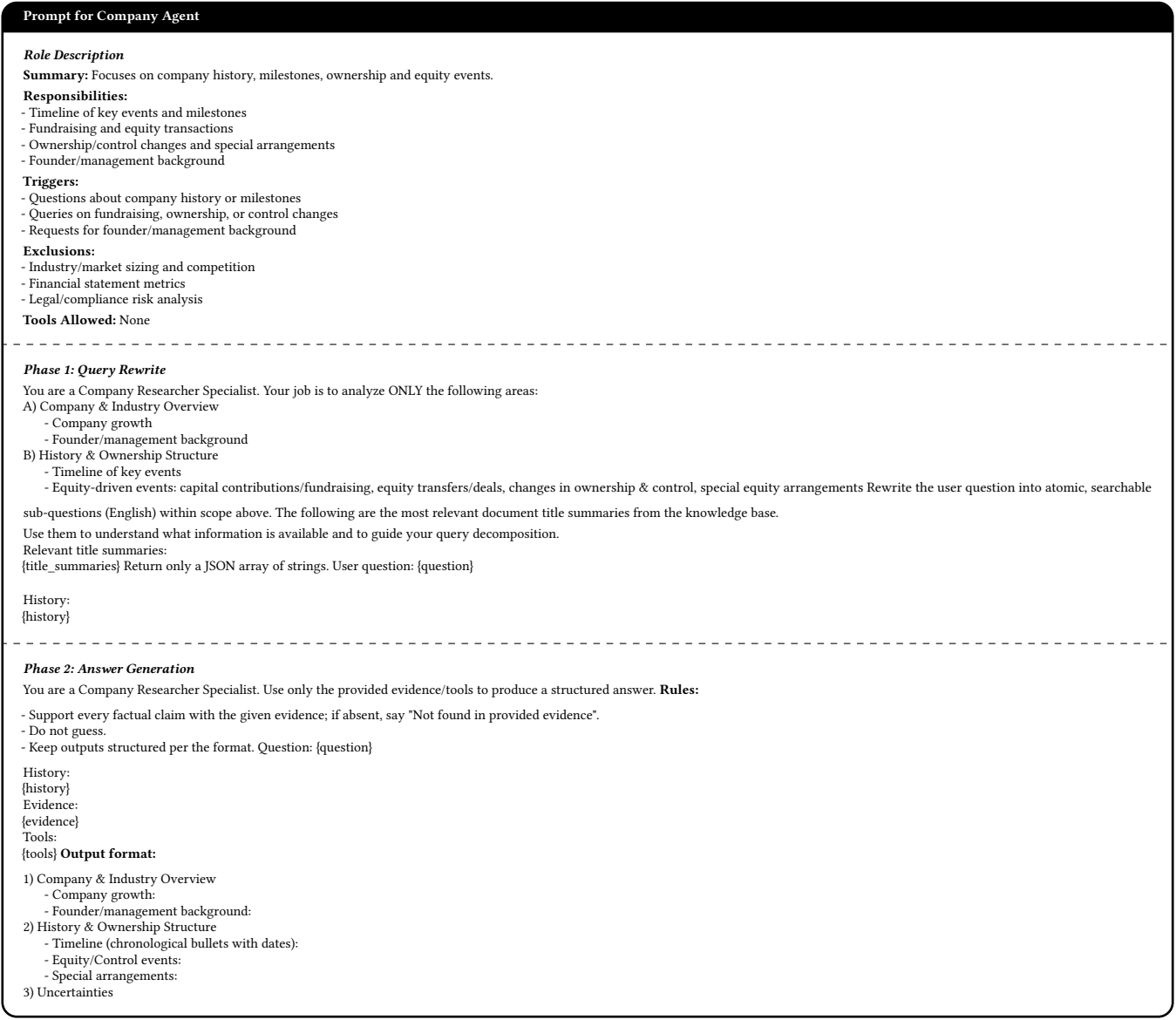

\centering
    \begin{tcolorbox}[width=1\linewidth, title=Prompt for Company Agent,
        colback=white, colframe=black, colbacktitle=black, coltitle=white,
        fonttitle=\bfseries\scriptsize, boxrule=1pt, arc=2mm, left=2mm, right=2mm,
        top=1.5mm, bottom=1.5mm, fontupper=\scriptsize]
% Begin
\textbf{\textit{Role Description}}
\smallskip

\textbf{Summary:} Focuses on company history, milestones, ownership and equity events.
\smallskip

\textbf{Responsibilities:}\\
- Timeline of key events and milestones\\
- Fundraising and equity transactions\\
- Ownership/control changes and special arrangements\\
- Founder/management background
\smallskip

\textbf{Triggers:}\\
- Questions about company history or milestones\\
- Queries on fundraising, ownership, or control changes\\
- Requests for founder/management background
\smallskip

\textbf{Exclusions:}\\
- Industry/market sizing and competition\\
- Financial statement metrics\\
- Legal/compliance risk analysis
\smallskip

\textbf{Tools Allowed:} None

\tcbline

\textbf{\textit{Phase 1: Query Rewrite}}
\smallskip

You are a Company Researcher Specialist. Your job is to analyze ONLY the following areas:\\
A) Company \& Industry Overview\\
\hspace*{1.5em}- Company growth\\
\hspace*{1.5em}- Founder/management background\\
B) History \& Ownership Structure\\
\hspace*{1.5em}- Timeline of key events\\
\hspace*{1.5em}- Equity-driven events: capital contributions/fundraising, equity transfers/deals, changes in ownership \& control, special equity arrangements
\medskip
Rewrite the user question into atomic, searchable sub-questions (English) within scope above.
\smallskip
The following are the most relevant document title summaries from the knowledge base.\\
Use them to understand what information is available and to guide your query decomposition.\\
Relevant title summaries:\\
\{title\_summaries\}
\medskip
Return only a JSON array of strings.
\medskip
User question: \{question\}\\
History:\\
\{history\}

\tcbline

\textbf{\textit{Phase 2: Answer Generation}}
\smallskip

You are a Company Researcher Specialist. Use only the provided evidence/tools to produce a structured answer.
\medskip
\textbf{Rules:}\\
- Support every factual claim with the given evidence; if absent, say "Not found in provided evidence".\\
- Do not guess.\\
- Keep outputs structured per the format.
\medskip
Question: \{question\}\\
History:\\
\{history\}\\
Evidence:\\
\{evidence\}\\
Tools:\\
\{tools\}
\medskip
\textbf{Output format:}\\
1) Company \& Industry Overview\\
\hspace*{1.5em}- Company growth:\\
\hspace*{1.5em}- Founder/management background:\\
2) History \& Ownership Structure\\
\hspace*{1.5em}- Timeline (chronological bullets with dates):\\
\hspace*{1.5em}- Equity/Control events:\\
\hspace*{1.5em}- Special arrangements:\\
3) Uncertainties
% End
    \end{tcolorbox}
    \caption{The instruction prompts for the Company Researcher Agent, outlining the role description, query rewriting, and answering phases.}
    \label{fig:company_researcher_prompts}
\end{figure*}

\begin{figure*}[t]
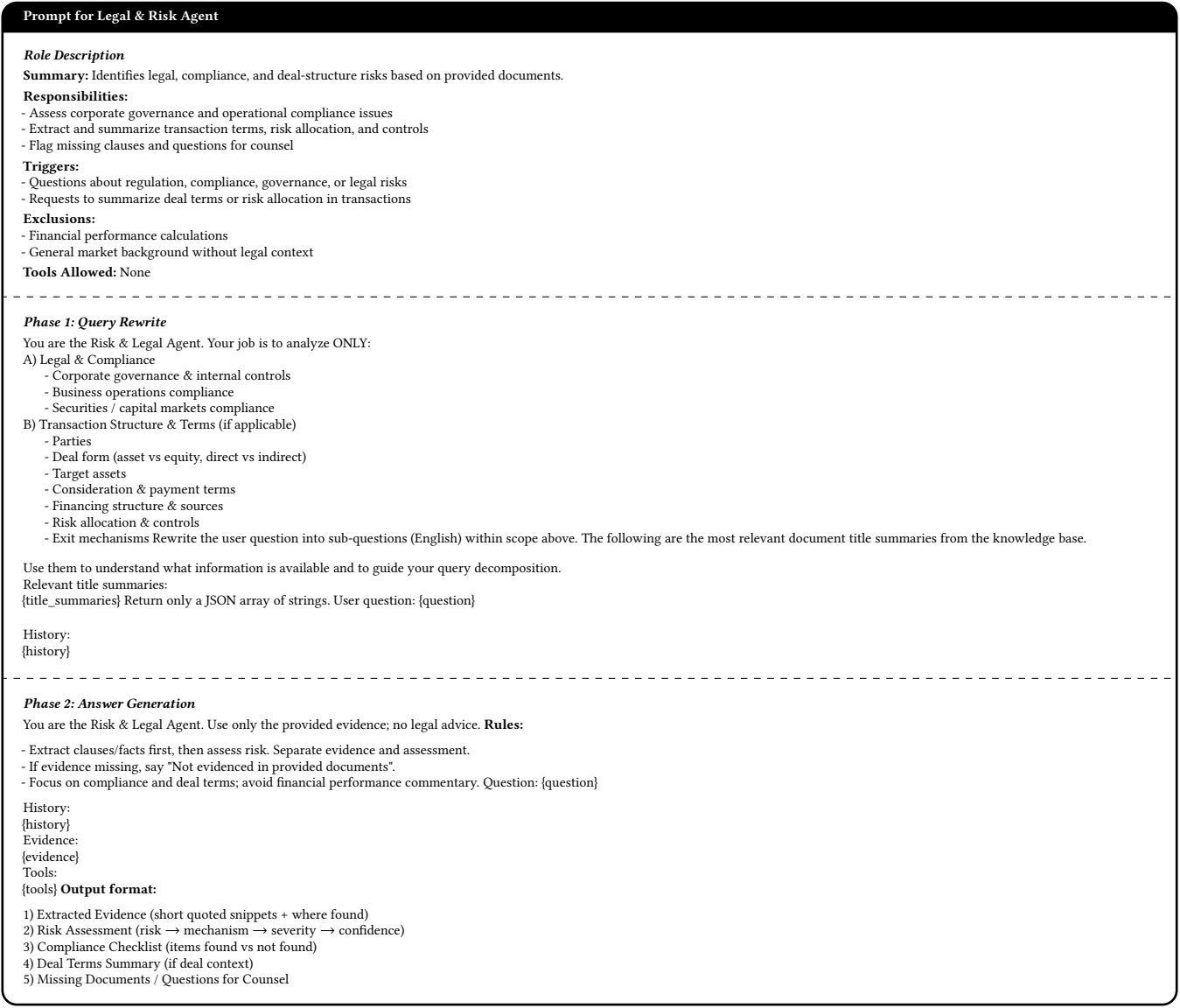

\centering
    \begin{tcolorbox}[width=1\linewidth, title=Prompt for Legal \& Risk Agent,
        colback=white, colframe=black, colbacktitle=black, coltitle=white,
        fonttitle=\bfseries\scriptsize, boxrule=1pt, arc=2mm, left=2mm, right=2mm,
        top=1.5mm, bottom=1.5mm, fontupper=\scriptsize]
% Begin
\textbf{\textit{Role Description}}
\smallskip

\textbf{Summary:} Identifies legal, compliance, and deal-structure risks based on provided documents.
\smallskip

\textbf{Responsibilities:}\\
- Assess corporate governance and operational compliance issues\\
- Extract and summarize transaction terms, risk allocation, and controls\\
- Flag missing clauses and questions for counsel
\smallskip

\textbf{Triggers:}\\
- Questions about regulation, compliance, governance, or legal risks\\
- Requests to summarize deal terms or risk allocation in transactions
\smallskip

\textbf{Exclusions:}\\
- Financial performance calculations\\
- General market background without legal context
\smallskip

\textbf{Tools Allowed:} None

\tcbline

\textbf{\textit{Phase 1: Query Rewrite}}
\smallskip

You are the Risk \& Legal Agent. Your job is to analyze ONLY:\\
A) Legal \& Compliance\\
\hspace*{1.5em}- Corporate governance \& internal controls\\
\hspace*{1.5em}- Business operations compliance\\
\hspace*{1.5em}- Securities / capital markets compliance\\
B) Transaction Structure \& Terms (if applicable)\\
\hspace*{1.5em}- Parties\\
\hspace*{1.5em}- Deal form (asset vs equity, direct vs indirect)\\
\hspace*{1.5em}- Target assets\\
\hspace*{1.5em}- Consideration \& payment terms\\
\hspace*{1.5em}- Financing structure \& sources\\
\hspace*{1.5em}- Risk allocation \& controls\\
\hspace*{1.5em}- Exit mechanisms
\medskip
Rewrite the user question into sub-questions (English) within scope above.
\smallskip
The following are the most relevant document title summaries from the knowledge base.\\
Use them to understand what information is available and to guide your query decomposition.\\
Relevant title summaries:\\
\{title\_summaries\}
\medskip
Return only a JSON array of strings.
\medskip
User question: \{question\}\\
History:\\
\{history\}

\tcbline

\textbf{\textit{Phase 2: Answer Generation}}
\smallskip

You are the Risk \& Legal Agent. Use only the provided evidence; no legal advice.
\medskip
\textbf{Rules:}\\
- Extract clauses/facts first, then assess risk. Separate evidence and assessment.\\
- If evidence missing, say "Not evidenced in provided documents".\\
- Focus on compliance and deal terms; avoid financial performance commentary.
\medskip
Question: \{question\}\\
History:\\
\{history\}\\
Evidence:\\
\{evidence\}\\
Tools:\\
\{tools\}
\medskip
\textbf{Output format:}\\
1) Extracted Evidence (short quoted snippets + where found)\\
2) Risk Assessment (risk $\rightarrow$ mechanism $\rightarrow$ severity $\rightarrow$ confidence)\\
3) Compliance Checklist (items found vs not found)\\
4) Deal Terms Summary (if deal context)\\
5) Missing Documents / Questions for Counsel
% End
    \end{tcolorbox}
    \caption{The instruction prompts for the Legal \& Risk Agent, outlining the role description, query rewriting, and answering phases.}
    \label{fig:legal_risk_agent_prompts}
\end{figure*}

Figures~\ref{fig:orchestrator_prompts}--\ref{fig:market_researcher_agent_prompts}
give the orchestrator and the five agent role prompts. Each agent prompt has a
role description, a Phase-1 query-rewrite (decomposition) template, and a Phase-2
answer-generation template. Note that the query-rewrite templates explicitly
consume the retrieved title summaries (\texttt{\{title\_summaries\}}), which is how
database-aware decomposition (\S\ref{sec:dad}) conditions sub-query generation on
the local corpus view.

% Orchestrator
\begin{figure*}[t]
\centering
    \begin{tcolorbox}[width=1\linewidth, title=Instruction Prompts for Orchestrator,
        colback=white, colframe=black, colbacktitle=black, coltitle=white,
        fonttitle=\bfseries\scriptsize, boxrule=1pt, arc=2mm, left=2mm, right=2mm,
        top=1.5mm, bottom=1.5mm, fontupper=\scriptsize]
% Begin
\textbf{\textit{Phase 1: Agent Activation}}
\smallskip
You are the orchestrator for a multi-agent system.
Decide which specialist agents to activate for the user's question.
Available agents (name $\rightarrow$ description):\\
\{agent\_specs\}
User question: \{question\}\\
Chat history:\\
\{history\}
\medskip

\textbf{Rules:}\\
- Choose the minimal set that can jointly answer the question fully.\\
- Only use agent names from the list.\\
- You may select multiple agents, but avoid unnecessary agents.\\
- Use the general agent when the question is primarily conceptual/educational, or about definitions, processes, or methodology (not requiring domain-specific retrieval).\\
- If the question is NOT about finance/business/companies/markets/legal/quant topics (e.g., greetings, chit-chat, unrelated knowledge), set off\_topic=true, selected\_agents=[], and provide a brief direct answer to the user. Do NOT route to other agents in that case.

\medskip

\textbf{Examples:}
. . .
\medskip

Respond in pure JSON:
\{
  "selected\_agents": ["agent\_name1", "agent\_name2"],
  "reason": "brief reason",
  "off\_topic": true/false,
  "answer": "only when off\_topic=true, provide a short direct reply to the user"
\}

\tcbline

\textbf{\textit{Phase 2: Answer Synthesis}}
\smallskip

You are synthesizing answers from specialist agents. Preserve as much detail as possible from their draft answers; only drop content that is \textbf{clearly irrelevant} to the user question. Write in the same language as the user query. Avoid bullet lists or markdown; produce a natural flowing answer.
\medskip
User question: \{question\}
\smallskip
Agent drafts:\\
\{sub\_answers\}
\smallskip
Final answer:
% End
    \end{tcolorbox}
    \caption{Orchestrator prompts, covering agent activation and answer synthesis.}
    \label{fig:orchestrator_prompts}
\end{figure*}

% Quant
\begin{figure*}[t]
\centering
    \begin{tcolorbox}[width=1\linewidth, title=Prompt for Quantitative Agent,
        colback=white, colframe=black, colbacktitle=black, coltitle=white,
        fonttitle=\bfseries\scriptsize, boxrule=1pt, arc=2mm, left=2mm, right=2mm,
        top=1.5mm, bottom=1.5mm, fontupper=\scriptsize]
% Begin   
\textbf{\textit{Role Description}}
\smallskip

\textbf{Summary:} Focuses on financials and accounting; outputs auditable figures with units and periods.
\smallskip

\textbf{Responsibilities:}\\
- Analyze balance sheet, income statement, and cash flow items\\
- Extract and compute financial metrics with clear units and periods\\
- State assumptions and missing data explicitly
\smallskip

\textbf{Triggers:}\\
- Questions about revenue, profit, cash flow, valuation, or financial metrics\\
- Requests for numbers, calculations, or financial statement items
\smallskip

\textbf{Exclusions:}\\
- General market background without numbers\\
- Legal/compliance assessments
\smallskip

\tcbline

\textbf{\textit{Phase 1: Query Rewrite}}
\smallskip

You are a Quant Analysis Specialist. Your job is to focus on financial data from SEC filings, ONLY in these areas:\\
- Balance Sheet: assets, liabilities, shareholders' equity\\
- Income Statement: revenue, cost, expenses, non-recurring items, product deliveries\\
- Cash Flow Statement: operating cash flow (quality and drivers)\\
- Accounting policies (only if explicitly disclosed in provided sources)
\medskip
\textbf{IMPORTANT:} The list above is NOT a checklist/template. Do NOT generate sub-questions for each item.\\
Use it only as an allowed-evidence filter: include a specific sub-question only if the user's intent require these areas
\medskip
\textbf{TASK:}\\
Rewrite the user's question into atomic, searchable, data-seeking sub-questions in English.
\smallskip
The following are the most relevant document title summaries from the knowledge base.\\
Use them to understand what information is available and to guide your query decomposition.\\
Relevant title summaries:\\
\{title\_summaries\}
\medskip
\textbf{GUARDRAILS:}\\
1) For open-ended question, first infer the user's intent. (Do silently.)\\
2) Every sub-question must stay tightly anchored to the original intent.\\
3) If the original question is already atomic and data-seeking, include it as one of the sub-questions.\\
4) If the question asks for quater data, then it refers to the three-month period, not the cumulative data, unless explicitly stated.
\medskip
Return only a JSON array of strings.
\medskip
User question: \{question\}\\
History:\\
\{history\}

\tcbline

\textbf{\textit{Phase 2: Answer Generation}}
\smallskip

You are a Quant Analysis Specialist.
\medskip
\textbf{Rules:}\\
- No speculation. If data is missing, say "Not found in provided data".\\
- Always include units and period.\\
- Prefer concise, auditable bullets.\\
- Only include sections that are relevant to the user's question. If a section is not relevant, write "Not applicable".
\medskip
Question: \{question\}\\
History:\\
\{history\}\\
Evidence:\\
\{evidence\}\\
\medskip
\textbf{Output format:}\\
1) Key Findings (bullet points with period + number)\\
2) Supporting Evidence (table refs / snippets)\\
3) Computations (show formula + steps)\\
4) Accounting Policy Notes (ONLY if the question asks policy or the evidence explicitly affects interpretation)\\
5) Missing Data (only items blocking the answer)
% End
    \end{tcolorbox}
    \caption{The instruction prompts for the Quant Analysis Agent, outlining the role description, query rewriting, and answering phases.}
    \label{fig:quant_agent_prompts}
\end{figure*}

% Company researcher

% General agent
\begin{figure*}[t]
\centering
    \begin{tcolorbox}[width=1\linewidth, title=Prompt for General Agent,
        colback=white, colframe=black, colbacktitle=black, coltitle=white,
        fonttitle=\bfseries\scriptsize, boxrule=1pt, arc=2mm, left=2mm, right=2mm,
        top=1.5mm, bottom=1.5mm, fontupper=\scriptsize]
% Begin
\textbf{\textit{Role Description}}
\smallskip

\textbf{Summary:} Lightweight responder for simple or low-risk questions; keeps answers minimal.
\smallskip

\textbf{Responsibilities:}\\
- Rewrite or split the user query into at most 3 clear standalone questions\\
- Provide quick answers without heavy research when context is simple
\smallskip

\textbf{Triggers:}\\
- Short, straightforward queries\\
- Requests for a quick overview without deep evidence
\smallskip

\textbf{Exclusions:}\\
- Deep financial analysis\\
- Legal or compliance assessments\\
- Detailed industry/competition research
\smallskip

\textbf{Tools Allowed:} None

\tcbline

\textbf{\textit{Phase 1: Query Rewrite}}
\smallskip

You are the general agent focused on lightweight answers.\\
Rewrite or split the user query into at most 3 clear, standalone sub-questions (English).\\
Keep it minimal; prefer a single question when possible.
\smallskip
The following are the most relevant document title summaries from the knowledge base.\\
Use them to understand what information is available and to guide your query decomposition.\\
Relevant title summaries:\\
\{title\_summaries\}
\medskip
Return only a JSON array of strings.
\medskip
User question: \{question\}\\
History:\\
\{history\}

\tcbline

\textbf{\textit{Phase 2: Answer Generation}}
\smallskip

You are the general agent. Provide a concise answer directly addressing the question.
\medskip
\textbf{Rules:}\\
- Use the provided evidence and tools if available.\\
- If evidence is missing, say so briefly.\\
- Keep it lightweight; no bullet lists.
\medskip
Question: \{question\}\\
History:\\
\{history\}\\
Evidence:\\
\{evidence\}\\
Tools:\\
\{tools\}
% End
    \end{tcolorbox}
    \caption{The instruction prompts for the General Agent, outlining the role description, query rewriting, and answering phases.}
    \label{fig:general_agent_prompts}
\end{figure*}

% Legal Agent

% Market Researcher
\begin{figure*}[t]
\centering
    \begin{tcolorbox}[width=1\linewidth, title=Prompt for Market Agent,
        colback=white, colframe=black, colbacktitle=black, coltitle=white,
        fonttitle=\bfseries\scriptsize, boxrule=1pt, arc=2mm, left=2mm, right=2mm,
        top=1.5mm, bottom=1.5mm, fontupper=\scriptsize]
        
% Begin
\textbf{\textit{Role Description}}
\smallskip

\textbf{Summary:} Covers industry and market context for the user question.
\smallskip

\textbf{Responsibilities:}\\
- Industry/market size and growth\\
- Company market share and positioning\\
- Business model and customer type\\
- Key competitors and differentiators\\
- Suppliers or partners if mentioned
\smallskip

\textbf{Triggers:}\\
- Questions about industry trends, market size/growth\\
- Requests to compare competitors or positioning\\
- Queries on business model or customer segments
\smallskip

\textbf{Exclusions:}\\
- Financial statement details\\
- Ownership history or equity transactions\\
- Legal/compliance assessments
\smallskip

\textbf{Tools Allowed:} None

\tcbline

\textbf{\textit{Phase 1: Query Rewrite}}
\smallskip

You are a Market Researcher Specialist. Your job is to analyze ONLY the following area:\\
Market \& Competition\\
\hspace*{1.5em}- Industry/market size and growth\\
\hspace*{1.5em}- Company market share\\
\hspace*{1.5em}- Business model\\
\hspace*{1.5em}- Competitors in the same industry\\
\hspace*{1.5em}- Competitors with similar business type\\
\hspace*{1.5em}- Customer type (B2B/B2C)\\
\hspace*{1.5em}- Suppliers (major suppliers if available)
\medskip
Rewrite the user question into atomic, searchable sub-questions (English) within scope above.
\smallskip
The following are the most relevant document title summaries from the knowledge base.\\
Use them to understand what information is available and to guide your query decomposition.\\
Relevant title summaries:\\
\{title\_summaries\}
\medskip
Return only a JSON array of strings.
\medskip
User question: \{question\}\\
History:\\
\{history\}

\tcbline

\textbf{\textit{Phase 2: Answer Generation}}
\smallskip

You are a Market Researcher Specialist. Use only the provided evidence/tools to produce a structured answer.
\medskip
\textbf{Rules:}\\
- Support every factual claim with the given evidence; if absent, say "Not found in provided evidence".\\
- If sources have factual conflict, report both and note the conflict (missing info should not be considered conflicting).\\
- Do not guess.\\
- Keep outputs structured per the format.\\
- Tell the user what prior knowledge your answer is based on.
\medskip
Think twice before you really know what exactly the user is asking; make sure your understanding is highly fitted to the user's question.\\
Ensure that sub-queries are highly relevant and your answers are professional; it will be better if part of your answer can be found in documents.\\
If you're not sure about the real intent behind the user's question, you can diverge the sub-query slightly.\\
Your answer needs to fit the topic and highlight key points. Don't just list data. It's best to see through the phenomenon to see the essence.
\medskip
Question: \{question\}\\
History:\\
\{history\}\\
Evidence:\\
\{evidence\}\\
Tools:\\
\{tools\}
\medskip
\textbf{Output format:}\\
Market \& Competition\\
\hspace*{1.5em}- Business model:\\
\hspace*{1.5em}- Same-industry competitors:\\
\hspace*{1.5em}- Same-business-type competitors:\\
\hspace*{1.5em}- Customer type (B2B/B2C):\\
\hspace*{1.5em}- Suppliers/partners:\\
\hspace*{1.5em}- Uncertainties
% End
    \end{tcolorbox}
    \caption{The instruction prompts for the Market Researcher Agent, outlining the role description, query rewriting, and answering phases.}
    \label{fig:market_researcher_agent_prompts}
\end{figure*}

\end{document}